\documentclass[a4paper,fleqn,usenatbib]{mnras}

\usepackage{newtxtext,newtxmath}
\usepackage[T1]{fontenc}
\usepackage{ae,aecompl}

\usepackage{color}

\usepackage{graphicx}	% Including figure files
\usepackage{amsmath}	% Advanced maths commands
\usepackage{amssymb}	% Extra maths symbols
\usepackage{epstopdf}

\newcommand{\rme}{\,\mathrm{e}}
\newcommand{\rmd}{\,\mathrm{d}}

\newcommand{\del}{\partial}

\title[Strong magnetic fields and g-mode spacings]{Effect of a strong magnetic field on gravity-mode period spacings in red giant stars}

\author[S.~T.~Loi]{
Shyeh Tjing Loi\thanks{E-mail: stl36@cam.ac.uk}
\\
Department of Applied Mathematics and Theoretical Physics, University of Cambridge, Centre for Mathematical Sciences, Wilberforce Road, Cambridge CB3 0WA, UK
}

\date{Accepted XXX. Received YYY; in original form ZZZ}

\pubyear{2020}

\begin{document}
\label{firstpage}
\pagerange{\pageref{firstpage}--\pageref{lastpage}}
\maketitle

\begin{abstract}
When a star evolves into a red giant, the enhanced coupling between core-based gravity modes and envelope-based pressure modes forms mixed modes, allowing its deep interior to be probed by asteroseismology. The ability to obtain information about stellar interiors is important for constraining theories of stellar structure and evolution, for which the origin of various discrepancies between prediction and observation are still under debate. Ongoing speculation surrounds the possibility that some red giant stars may harbour strong (dynamically significant) magnetic fields in their cores, but interpretation of the observational data remains controversial. In part, this is tied to shortfalls in our understanding of the effects of strong fields on the seismic properties of gravity modes, which lies beyond the regime of standard perturbative methods. Here we seek to investigate the effect of a strong magnetic field on the asymptotic period spacings of gravity modes. We use a Hamiltonian ray approach to measure the volume of phase space occupied by mode-forming rays, this being roughly proportional to the average density of modes (number of modes per unit frequency interval). A strong field appears to systematically increase this by about 10\%, which predicts a $\sim$10\% smaller period spacing. Evidence of near integrability in the ray dynamics hints that the gravity-mode spectrum may still exhibit pseudo-regularities under a strong field.
\end{abstract}


%%%%%%%%%%%%%%%%%%%%%%%%%%%%%%%%%%%%%%%%%%%%%%%%%%

%%%%%%%%%%%%%%%%% BODY OF PAPER %%%%%%%%%%%%%%%%%%

\section{Introduction}\label{sec:intro}

The nature of magnetic fields below the surfaces of stars remain elusive to direct observation. The strength and distribution of interior fields is likely to depend on their origin (for example, accretion during star formation, generation by an active dynamo, or a relic from a previous dynamo phase), but uncertainties surround many aspects of associated theories \citep{Braithwaite2017}. Most stars exhibit surface fields, and so while the existence of an interior field that is a continuation of the observed surface field is unquestioned, it is often difficult to infer much more. Alongside these ongoing puzzles, advances in the theory of stellar evolution and the increasing number of discrepancies between prediction and observation have called into question the role of magnetic fields, the speculation being that account of their effects may offer at least a partial resolution to various discrepancies \citep{Mathis2013}. These issues include angular momentum transport, where the observed slow rotation rates of evolved stellar cores and stellar remnants necessitate transport efficiencies greater than those provided by known hydrodynamic processes. Magnetic fields with their associated Maxwell stresses naturally provide a means of assisting this, but the exact mechanisms and their sufficiency remain under debate \citep[e.g.][]{Fuller2019, Hartogh2020}. In addition, explaining photospheric chemical abundances in evolved stars requires mixing processes operating in radiative zones beyond those of standard models \citep{Mathis2005}. It is thought that this might also be achievable by magnetic means \citep{Busso2007, Charbonnel2007}.

Progress with these issues would benefit greatly from improved constraints of stellar interior properties. A powerful technique for probing the interiors of stars is asteroseismology, which involves measurements of the frequencies of global oscillation (through time-dependent brightness variations) combined with a suitable theoretical framework for interpretation. While the hydrodynamical aspects of the theory of stellar oscillations are well established, the inclusion of magnetic fields, particular those of dynamically significant strengths, represents an extension still under development. Improving our understanding of how magnetic fields affect the seismic spectrum will ultimately provide more clues about the nature of magnetism deep within stellar interiors, with broader impacts upon unsolved problems in stellar physics.

On a fundamental level, global modes of oscillation form from the constructive interference of waves. The structure of the resulting spectrum depends on the properties of the object supporting their propagation. Such concepts were originally used to understand the Schr\"{o}dinger equation and the quantisation of atomic spectra \citep{Keller1958}. Their applicability to general eigenvalue problems was expounded by \citet{Keller1960}, and appropriated by \citet{Gough1986} for the treatment of stellar oscillations. Eigenmode formation requires that the phase elapsed over any closed path $C_i$ is an integer multiple of $2\pi$, i.e.
\begin{align}
  \oint_{C_i} \mathbf{k} \cdot \rmd \mathbf{x} = 2\pi \left( n_i + \frac{\beta_i}{4} \right) \:,\quad n_i \in \mathbb{Z} \:, \label{eq:EBK}
\end{align}
where $i = 1, \cdots, \mathcal{N}$ index the $\mathcal{N}$ spatial degrees of freedom, $\mathbf{k}$ is the wavevector, $\mathbf{x}$ is the position vector, and $\beta_i$ are the Maslov indices associated with phase jumps at caustic surfaces. The integers $n_i$, which count the number of wavelengths around an orbit, relate to quantum numbers of the modes. Equation (\ref{eq:EBK}) is known as the Einstein-Brillouin-Keller (EBK) quantisation condition after its discoverers \citep{Einstein1917, Brillouin1926, Keller1958}.

The quantum numbers $n_i$ from EBK quantisation are closely related to conserved actions in the framework of Hamiltonian dynamics. Continuum mechanics is amenable to a Hamiltonian treatment when wavelengths are much shorter than the scales of background variation. This approach has been widely used to study wave propagation in space plasmas \citep{Haselgrove1955, Walker2004}, the Earth's interior and atmosphere \citep{Chian1994, Hasha2008, Lu2008}, and stellar interiors \citep{Lignieres2011, Pasek2013}. For an integrable system, i.e.~where as many symmetries as spatial degrees of freedom exist, a canonical transformation $(q_i, p_i) \to (Q_i, P_i)$ can be performed such that the Hamiltonian $H$ is independent of $Q_i$, for $i = 1, \cdots, \mathcal{N}$. Here $p_i$ and $P_i$ are the conjugate momenta associated with coordinates $q_i$ and $Q_i$. Since $\dot{Q}_i = \del H/\del P_i$ and $\dot{P}_i = -\del H/\del Q_i$, where dot indicates a time derivative, this implies that $P_i$ are constants of motion. They are referred to as \textit{actions} and can be computed as
\begin{align}
  P_i = \frac{1}{2\pi} \oint p_i \rmd q_i \:. \label{eq:action}
\end{align}
Notice that this has the same form as (\ref{eq:EBK}). Conserved quantities arise from symmetries present in the system \citep{Noether1918}. For example, time-translational symmetry (time independence) results in conservation of energy $E$ (equivalently, the wave frequency $\omega$), spherical symmetry leads to conservation of total angular momentum $L$, and rotational symmetry about the $z$-axis leads to conservation of the $z$-component of the angular momentum, $L_z$.

In a star whose backgrounds are spherically symmetric and not changing with time, the dynamics are integrable since three conserved quantities ($\omega$, $L$ and $L_z$) exist, equalling the number of spatial degrees of freedom, $\mathcal{N} = 3$. The entire mode spectrum is then described by three quantum numbers $n$, $\ell$, and $m$, known as the radial order, spherical degree, and azimuthal order respectively, and the resulting spectrum exhibits regular patterns. However, if rotation and/or a magnetic field is present, then spherical symmetry is broken, integrability is lost and the system transitions towards chaos. Modes associated with chaotic regions of phase space are not organised by a set of quantum numbers but rather exhibit irregular spectra, where the exact frequencies are exponentially sensitive to the stellar structure. A substantial body of work exists applying Hamiltonian dynamics to studying the behaviour of waves in stars where the symmetry has been broken by rapid rotation, both for acoustic waves \citep{Lignieres2008, Lignieres2009, Pasek2011} and gravity waves \citep{Prat2016, Prat2017, Prat2018}, and including magnetic fields \citep{Valade2018}. These works have established that wave chaos emerges under conditions of broken symmetry, and that the transition occurs gradually in a manner consistent with the Kolmogorov-Arnold-Moser theorem \citep{Kolmogorov1954, Moser1962, Arnold1963}, beginning in small regions of phase space that expand as the symmetry becomes more heavily broken. Notably, many of these works find that even at high rotation rates, there still exist parts of phase space where the dynamics are nearly integrable and EBK quantisation can be applied to obtain asymptotic formulae for the mode frequencies.

A useful quantity for describing eigenmode spectra is the density of modes. An analytic formula for the average density of modes was derived by \citet{Weyl1912}, the leading order term of which dominates in the high-energy limit and can be obtained from general principles if one considers each mode to occupy a fixed phase-space volume of $(2\pi)^\mathcal{N}$. The number of modes $n_W$ possessing energies below a certain value $E$ corresponds to the phase space volume associated with $H < E$, divided by $(2\pi)^\mathcal{N}$:
\begin{align}
  n_W(E) \approx \frac{1}{(2\pi)^\mathcal{N}}\int_{H < E} \rmd^\mathcal{N} \mathbf{k} \rmd^\mathcal{N} \mathbf{x} \:. \label{eq:nWeyl}
\end{align}
The average density of modes (number of modes per unit interval of energy) is then given by $\rmd n_W/\rmd E$. When applied to wave problems in fluid mechanics, energy $E$ is replaced by the wave frequency $\omega$ and $H$ is replaced by the dispersion relation of the wave mode in question.
In evolved stars, the two main types of wave mode present are (i) acoustic waves, restored by gas pressure and localised to the convective envelope, and (ii) gravity waves, restored by buoyancy and localised to the radiative core. These two propagation zones are known as the p-mode and g-mode cavities, respectively. The evanescent region separating these cavities is thin for low $\ell$, allowing for fluid motions to couple between them. This forms mixed modes with both acoustic- and gravity-mode character, easily observable due to large surface displacements yet highly sensitive to core properties, making them extremely useful for seismic diagnosis of the deep interior.

In this work, we consider any strong magnetic fields to be confined to the radiative core, where they might affect gravity wave propagation. We ignore any possible magnetic effects on the acoustic part of a mixed mode. Application of (\ref{eq:nWeyl}) to gravity modes of fixed $\ell$ and $m$ predicts the number of modes with frequencies greater than $\omega$ to be (see later in Section \ref{sec:volume_measurement})
\begin{align}
  n_W(\omega) \approx \frac{1}{\pi} \int_{r_1}^{r_2} \frac{\sqrt{\ell(\ell+1)}}{r} \left( \frac{N^2}{\omega^2} - 1 \right)^{1/2} \rmd r \:, \label{eq:ngrav}
\end{align}
where $r$ is the radial coordinate, $r_1$ and $r_2$ are the lower and upper turning points of the g-mode cavity, and $N$ is the buoyancy frequency.

For comparison, a Wentzel-Kramers-Brillouin-Jeffreys (WKBJ) treatment of the fluid equations yields the following asymptotic relation for the frequencies of g-modes of high radial order $n$, i.e.~in the limit where $N \gg \omega$ \citep{Tassoul1980, Gough2007}:
\begin{align}
  \int_{r_1}^{r_2} \frac{\sqrt{\ell(\ell+1)}}{r} \frac{N}{\omega} \rmd r = (n + \epsilon) \pi \:, \label{eq:n_WKB}
\end{align}
where $\epsilon$ is a phase offset sensitive to background properties in the vicinity of the evanescent zone. Taking the ``high-energy'' limit $N \gg \omega$, Equation (\ref{eq:ngrav}) reduces to the form in (\ref{eq:n_WKB}), and $n_W$ can be identified with the radial order $n$ (up to the offset $\epsilon \sim O(1) \ll n$). For constant or slowly-varying $\epsilon$, Equation (\ref{eq:n_WKB}) implies successive values of $\omega^{-1}$ to be uniformly spaced, leading to the prediction that g-modes of fixed $\ell$ and large $n$ should exhibit constant period spacing, this being proportional to $\int (N/r) \rmd r$.

Take note that although for pure g-modes, $n$ (measuring the number of radial crossings) and $n_W$ (measuring the volume of available phase space) are closely related, this connection weakens under a strong magnetic field. In this situation, modes may start to acquire irregular node patterns thus making $n$ difficult to define; $n_W$ then becomes the more reliable quantity for estimating the number of modes. We shall refer to $n_W$ as the ``effective'' radial order, in the sense that it acts as a counter for the number of modes, similar to radial order in the non-magnetic case. However, it is not directly related to numbers of radial crossings.

It can be seen that EBK quantisation, Hamiltonian conserved actions, the Weyl formula and the asymptotic theory of stellar oscillations are closely interlinked. This paper exploits the relationship between these concepts with the goal of quantifying the effect of a strong core magnetic field on the density of gravity modes in red giant stars. To do this, we use the ray tracing approach of \citet{Loi2020} to compute the propagation of magneto-gravity waves in an idealised model of a red giant core harbouring a magnetic field, and empirically measure the phase space volume occupied by periodic rays. By observing how this volume varies with frequency, we obtain an estimate of the density of modes $\rmd n_W/\rmd \omega$ at different field strengths. To first approximation, a study of the effects of a core magnetic field on the properties of core-propagating modes can afford to neglect coupling into the envelope. We simplify the problem by focusing solely on wave propagation within the g-mode cavity, which is assumed to wholly contain any regions of strong field. We also neglect rotational effects.

This investigation is more broadly motivated by the suggestion that strong magnetic fields may exist in the cores of red giant stars \citep{Fuller2015, Stello2016_mn2e}, which has been met with controversy, notably in the light of analyses which found g-mode asymptotic period spacings not to vary significantly across the red giant population \citep{Mosser2017}. The presumption that strong magnetic fields ought to affect g-mode spacings stems from the idea that the size of the propagation cavity could be significantly modified by a strong field, which would occur if gravity wave propagation is disrupted. However, recent work by \citet{Loi2020} suggested the existence of propagation paths where gravity waves do not significantly feel the Lorentz force and so remain undisrupted, and that associated modes might exhibit seismic properties similar to those in unmagnetised stars. We seek to follow up on this aspect here.

This paper is structured as follows. In Section \ref{sec:models} we introduce the stellar/magnetic field models and parameters. In Section \ref{sec:methods} we describe the ray tracing method and the approach to measuring the phase space volumes. We present results in Section \ref{sec:results}, and discuss broader impacts in Section \ref{sec:discussion}. Finally, we conclude in Section \ref{sec:summary}.

\section{Models \& parameters}\label{sec:models}
Here we outline the choice of functions to model the radiative core and embedded magnetic field. We use an idealised analytic model of a red giant core, with parameters chosen to emulate the $2\,M_\odot$ red giant model used by \citet{Loi2020}, therein `Model B'. In that work, the stellar model was generated using the publicly-available stellar evolutionary code `Modules for Experiments in Stellar Astrophysics' (\textsc{mesa}), but our choice here to adopt an analytic model rather than a \textsc{mesa}-generated one stems from the desire to reduce the numerical errors associated with integrating the ray tracing equations on a non-smooth background. The non-smoothness arises from the grid-based nature of the \textsc{mesa} output, where background quantities are defined at only a finite set of points necessitating interpolation to obtain intermediate values. The use of a continuously defined, infinitely differentiable analytic model circumvents these issues.

\subsection{Stellar model}
The differential equations governing the propagation of magneto-gravity waves (see later in Section \ref{sec:raytracing}) require only a small number of hydrodynamic background quantities to be defined, namely the buoyancy frequency $N$ and its spatial gradient. In principle, the magnetic field only needs to be defined via the Alfv\'{e}n velocity profile $\mathbf{v}_A$, but given the approach here to constructing the magnetic field, which defines a scalar function $\Psi$ relating directly to the magnetic field $\mathbf{B}$ rather than $\mathbf{v}_A$ (see later in Section \ref{sec:magneticfield}), it is necessary to also construct the mass density profile $\rho$, since $\mathbf{v}_A \propto \mathbf{B}/\sqrt{\rho}$.

\begin{figure}
  \centering
  \includegraphics[width=\columnwidth]{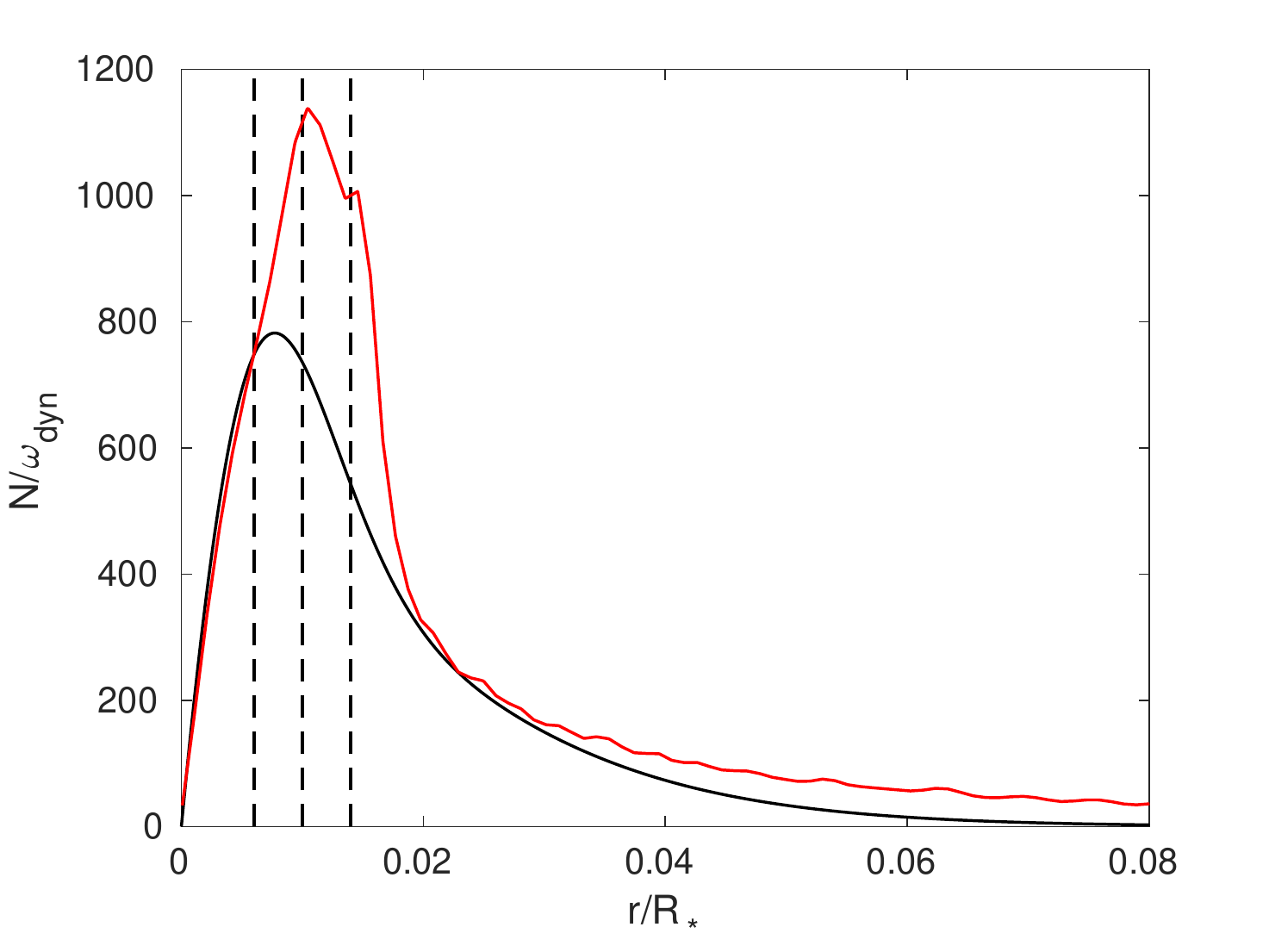}
  \includegraphics[width=\columnwidth]{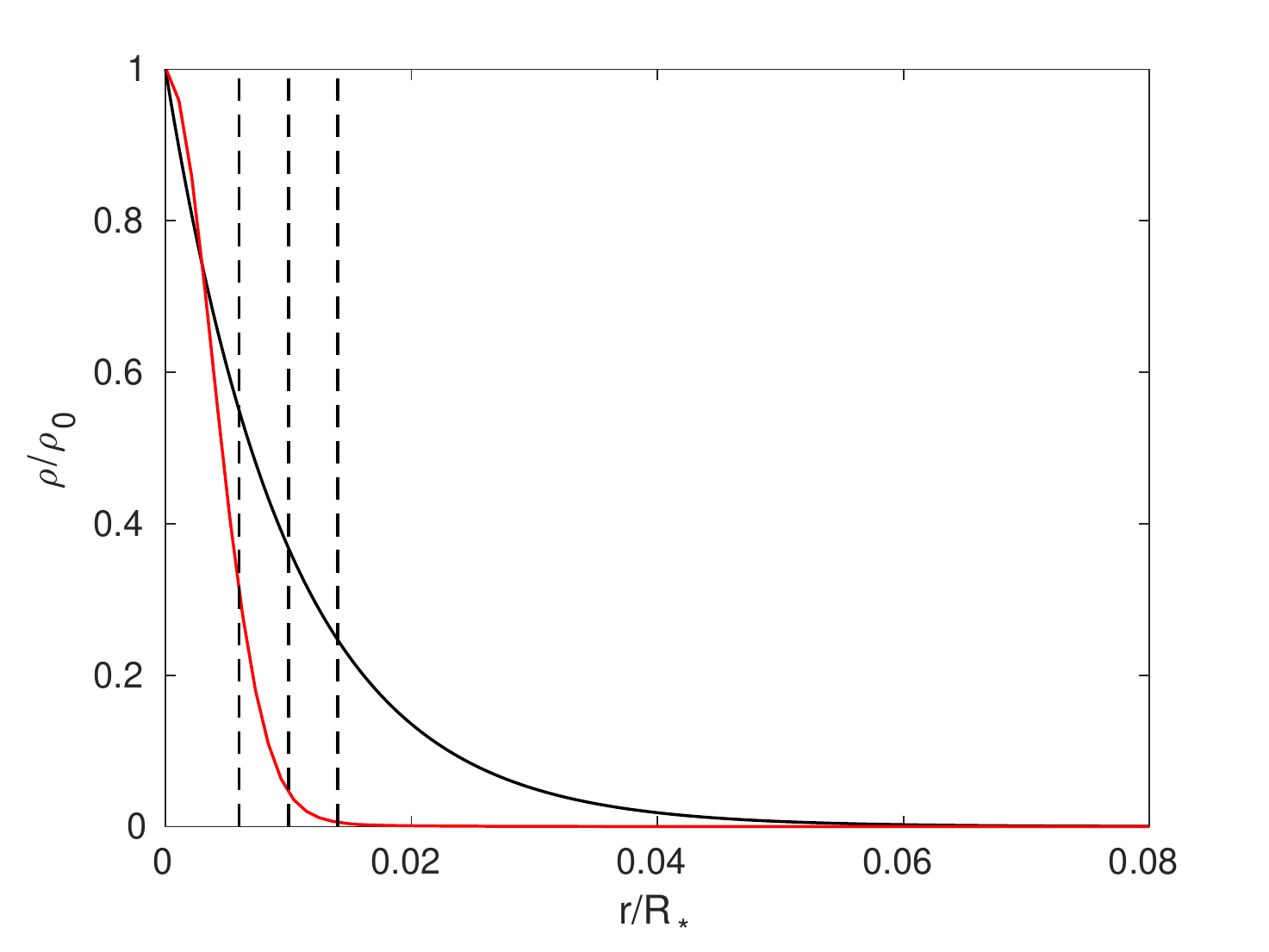}
  \caption{Top: Buoyancy frequency profile defined in Equation (\ref{eq:N}), with $N_0 = 1000$, $\sigma = 0.01$ (black). A \textsc{mesa} model of a red giant is shown for comparison (red). Vertical dashed lines indicate the three trial values of the field radius $R_f$, these being 0.006, 0.01 and 0.014\,$R_*$. The field is assumed to extend all the way to the centre and vanish for $r > R_f$. Bottom: Likewise for the density profile defined in Equation (\ref{eq:rho}).}
  \label{fig:Nprofile}
\end{figure}

Here we make the assumption that all hydrodynamic backgrounds are spherically symmetric, so $N$ and $\rho$ depend only on $r$. We adopt the functional forms
\begin{align}
  N(r) &= \frac{N_0 r}{\sigma} \left( \rme^{-r/\sigma} + \rme^{-r^2/\sigma^2} \right) \:, \label{eq:N} \\
  \rho(r) &= \rho_0 \rme^{-r/\sigma} \:, \label{eq:rho} 
\end{align}
where $N_0$, $\sigma$ and $\rho_0$ are arbitrary constant parameters. For this work we choose $N_0 = 1000$ and $\sigma = 0.01$, where length scales are expressed in units of the stellar radius $R_*$ and frequencies are in units of the dynamical frequency $\omega_\text{dyn} = \sqrt{GM_*/R_*^3}$, where $M_*$ is the stellar mass. The resulting buoyancy frequency and density profiles are shown in Figure \ref{fig:Nprofile}, with a 2\,$M_\odot$, 1.001\,Gyr-old \textsc{mesa} red giant model overplotted for comparison. It can be seen that the overall qualitative features of a g-mode cavity are reproduced: $N = 0$ at the centre, reaching a maximum some way from the centre and then decaying to larger $r$, while $\rho$ is maximum at the centre, decaying outwards on a length scale comparable to that of $N$. It is to be noted that the analytic density profile is cusp-like at the centre, rather than obeying $\rmd \rho/\rmd r = 0$ which would be expected in a real star. However, the finiteness of $\rmd \rho/\rmd r$ at $r = 0$ is not expected to adversely impact the ray dynamics, since curvature terms (which go as $1/r$) dominate over $\rmd \rho/\rmd r$ terms near the centre. Furthermore, the g-mode cavity does not extend to the origin, so no rays pass through $r = 0$.

Note that the gas pressure does not appear in the equations for magneto-gravity ray tracing, so from an algebraic point of view $\rho$ can be constructed independently of $N$. The shape of the $\rho$ profile influences the geometry of the magnetic field subsequently constructed (more detail in Section \ref{sec:magneticfield}); here it has been chosen to be a decaying exponential with the same characteristic decay length as $N$, since this qualitatively resembles the situation in a real star. Because of its sole purpose in constructing the field, the value of $\rho_0$ cannot be adjusted independently of the central field strength; since the latter is the parameter we wish to control, it can be assumed for simplicity that $\rho_0 = 1$.

\subsection{Magnetic field}\label{sec:magneticfield}
We employ a twisted-torus field model; configurations of this type are believed on the basis of analytical and numerical arguments to be a realistic description of a magnetic equilibrium \citep{Tayler1973, Flowers1977, Braithwaite2006, Duez2010}. An analytic formula for such an equilibrium configuration was derived by \citet{Prendergast1956}, where the field components $\mathbf{B} = (B_r, B_\theta, B_\phi)$ in spherical polar coordinates $(r, \theta, \phi)$ are given in terms of the radial flux function $\Psi(r)$ by
\begin{align}
  B_r &= \frac{2}{r^2} \Psi(r) \cos \theta \:, \label{eq:Br} \\
  B_\theta &= -\frac{1}{r} \Psi'(r) \sin \theta \:, \label{eq:Bth} \\
  B_\phi &= -\frac{\lambda}{r} \Psi(r) \sin \theta \:, \label{eq:Bphi}
\end{align}
where
\begin{align}
  \Psi(r) &= \frac{\beta \lambda r}{j_1(\lambda R_f)} \left[ f_\lambda(r, R_f) \int_0^r \rho \xi^3 j_1(\lambda \xi) \rmd \xi \right. \nonumber \\
    &\qquad \left. + j_1(\lambda r) \int_r^{R_f} \rho \xi^3 f_\lambda(\xi, R_f) \rmd \xi \right] \:. \label{eq:Psi}
\end{align}
Here $R_f$ is the radial extent of the field, $\beta$ is an arbitrary scaling factor (controlling strength of the field), $j_1$ and $y_1$ are spherical Bessel functions, $f_\lambda(r_1, r_2) \equiv j_1(\lambda r_2) y_1(\lambda r_1) - j_1(\lambda r_1) y_1(\lambda r_2)$, and $\lambda$ is the root of
\begin{align}
  \int_0^{R_f} \rho \xi^3 j_1(\lambda \xi) \rmd \xi = 0 \:. \label{eq:lambda}
\end{align}
This corresponds to an axisymmetric twisted torus, where all three components of the field vanish smoothly at the boundary $r = R_f$.

\begin{figure}
  \centering
  \includegraphics[width=\columnwidth]{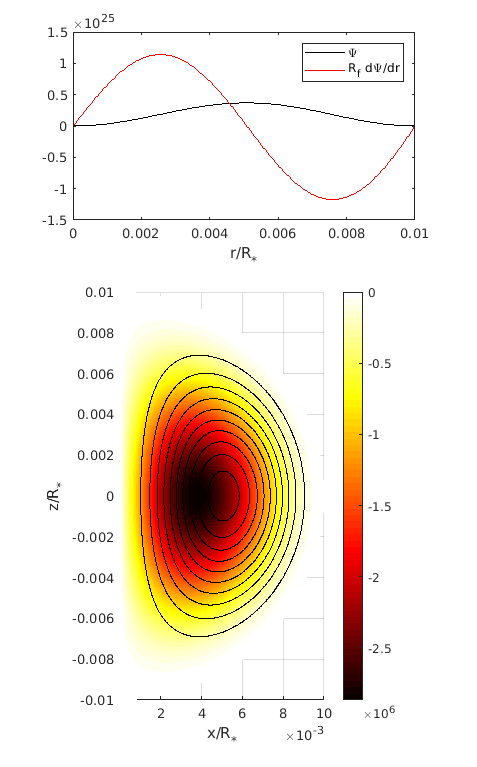}
  \caption{Twisted-torus magnetic field configuration used in this work, computed for $R_f = 0.01\,R_*$ and with the field strength scaled such that $v_{A,\text{cen}} = v_{A,\text{crit}}$. Near-identical solutions are obtained for the other values of $R_f$. For ease of interpretation the quantities are expressed in dimensional units, dimensionalised using $R_* = 6\,R_\odot$, $M_* = 2\,M_\odot$ and $\rho_0 = 5 \times 10^7\,$kg/m$^3$, in which case the central field strength is 4\,MG. Top: the radial flux function $\Psi$ and its derivative (scaled by $R_f$, both expressed in units of Mx), which define the field components according to Equations (\ref{eq:Br})--(\ref{eq:Bphi}). Bottom: a meridional half-section showing the toroidal component of the field in colour (units of G, where the negative sign indicates that this is directed out of the page) and the poloidal projections of the field lines as black contours. The smallest value of $\lambda$ satisfying (\ref{eq:lambda}) in this case was $\lambda = 605.02\,R_*^{-1}$.}
  \label{fig:Prendergast}
\end{figure}

If $\rho$ does not vary too rapidly compared to $R_f$, it should be possible to find value of $\lambda$ satisfying Equation (\ref{eq:lambda}). The factor $\xi^3 j_1(\lambda \xi)$ is a growing sinusoid, with oscillations occurring on the scale $\lambda^{-1}$. Since the integral of a growing sinusoid is likewise a growing sinusoid, possessing multiple zeroes (when regarded as a function of $R_f$), then appropriately increasing/decreasing $\lambda$ (rescaling the oscillatory pattern) would eventually enable one of these zeroes to coincide with some chosen $R_f$. As long as $\rho$ does not decrease too rapidly, the oscillatory nature of the integral is maintained which should enable a value of $\lambda$ to be found. Note that in general, because of this oscillatory behaviour, more than one root exists for any given $\rho$ and $R_f$. We selected the smallest value of $\lambda$ in each case, corresponding to the field solution with the largest scale of variation. Figure \ref{fig:Prendergast} shows the resulting field configuration for $R_f = 0.01$, for which the smallest root was $\lambda = 605.02$. However, if $\rho$ decreases steeply enough ($\sigma \ll R_f$), then the integrand becomes a decaying sinusoid and a root may not exist. For example, values of $\lambda$ can be found for $\sigma = 10^{-3}$ (smallest root $\lambda = 1.27 \times 10^4$), but not for $\sigma = 10^{-4}$, in the case where $R_f = 0.01$. Since we wanted to solve for a field the size of the core ($R_f \sim \sigma$) this was not a problem here, but it may be an issue for other applications. Note that Equations (\ref{eq:Br})--(\ref{eq:Psi}) apply to the region $r \leq R_f$. For $r > R_f$, we set $\Psi = 0$, i.e.~all field components were matched to the zero solution.

Conveniently, the choice of functional form for $\rho$ (a decaying exponential) means that the integrals in (\ref{eq:Psi}) can be done analytically. This circumvents the problem discussed at the start of this section surrounding the exacerbation of numerical errors by non-smooth backgrounds when performing ray tracing. The resulting expressions for $\Psi$ and its derivative can be found in Appendix \ref{sec:appendix}.
  
Magnetic fields can be expected to influence gravity wave propagation at strengths where the resonance condition (coincidence of frequencies and wavenumbers) between Alfv\'{e}n and gravity waves is satisfied. This can be used to define the critical Alfv\'{e}n speed
\begin{align}
  v_{A,\text{crit}} \sim \frac{r}{N} \frac{\omega^2}{\sqrt{\ell(\ell+1)}} \:,
\end{align}
where a ``strong'' field refers to one for which regions where $v_A > v_{A,\text{crit}}$ exist. For the stellar model examined, $r/N \sim 5 \times 10^{-6}$ in the core, which for typical wave parameters (see Section \ref{sec:waveparams}) implies that $v_{A,\text{crit}} \sim 2 \times 10^{-4} v_\text{dyn}$, where $v_\text{dyn} = \sqrt{GM_*/R_*}$ is the dynamical speed. For a red giant with $M_* = 2\,M_\odot$ and $R_* = 6\,R_\odot$, this translates to a critical Alfv\'{e}n speed of about 50\,m/s. Assuming a central density of $\rho_0 \approx 5 \times 10^7$\,kg/m$^3$ that might be found in such a star, this corresponds to a central field strength of $\sim$4\,MG. We adjusted the scaling factor $\beta$ so that the central field strength was equal to certain multiples of the critical strength\footnote{N.B.~the central value is nearly but not exactly equal to the maximum, which occurs slightly away from the centre and is about 4\% larger.}. For $R_f = 0.01$, we tested central Alfv\'{e}n speeds of $v_{A,\text{cen}} = 0.01, 0.1, 1$ and 10\,$v_{A,\text{crit}}$. Two additional values of $R_f$ were tested for comparison, these being $R_f = 0.006$ and 0.014, and in both cases we set $v_{A,\text{cen}} = v_{A,\text{crit}}$. These three values of $R_f$ are shown in Figure \ref{fig:Nprofile}. They were chosen to be around where one might expect the H-burning shell (boundary of the old convective core) to be present in a red giant, which is typically just outside the main maximum in $N$. However, the final extent of a magnetic field that may be left over from a previous core dynamo is not certain, as the relaxation process may be accompanied by reconnection, annihilation and/or a net inward or outward motion of magnetic flux. The three $R_f$ values were roughly placed in the vicinity of the maximum in $N$, differing by over a factor of two, to reflect this uncertainty.

\subsection{Wave parameters}\label{sec:waveparams}
In stars whose oscillations are driven stochastically by convective motions, the frequency of maximum power is strongly correlated with the dynamical frequency $\omega_\text{dyn} = \sqrt{GM_*/R_*^3}$, where $M_*$ is the stellar mass \citep{Bedding2013}. For red giants, this occurs near $10\,\omega_\text{dyn}$ \citep{Mosser2013}. Modes are often seen in a broad envelope about this maximum, whose width may span several times $\omega_\text{dyn}$. We tested five values of the wave frequency $\omega$, these being 8, 9, 10, 11 and 12\,$\omega_\text{dyn}$.

Since only low spherical degrees $\ell$ are observable for distant stars (due to geometric cancellation effects), we focused on a value of $\ell = 1$ for all values of $R_f$ and $v_{A,\text{cen}}$. In addition, for $R_f = 0.01\,R_*$ and $v_{A,\text{cen}} = v_{A,\text{crit}}$, we also tested $\ell = 2$ and 3.

\begin{figure}
  \centering
  \includegraphics[width=\columnwidth]{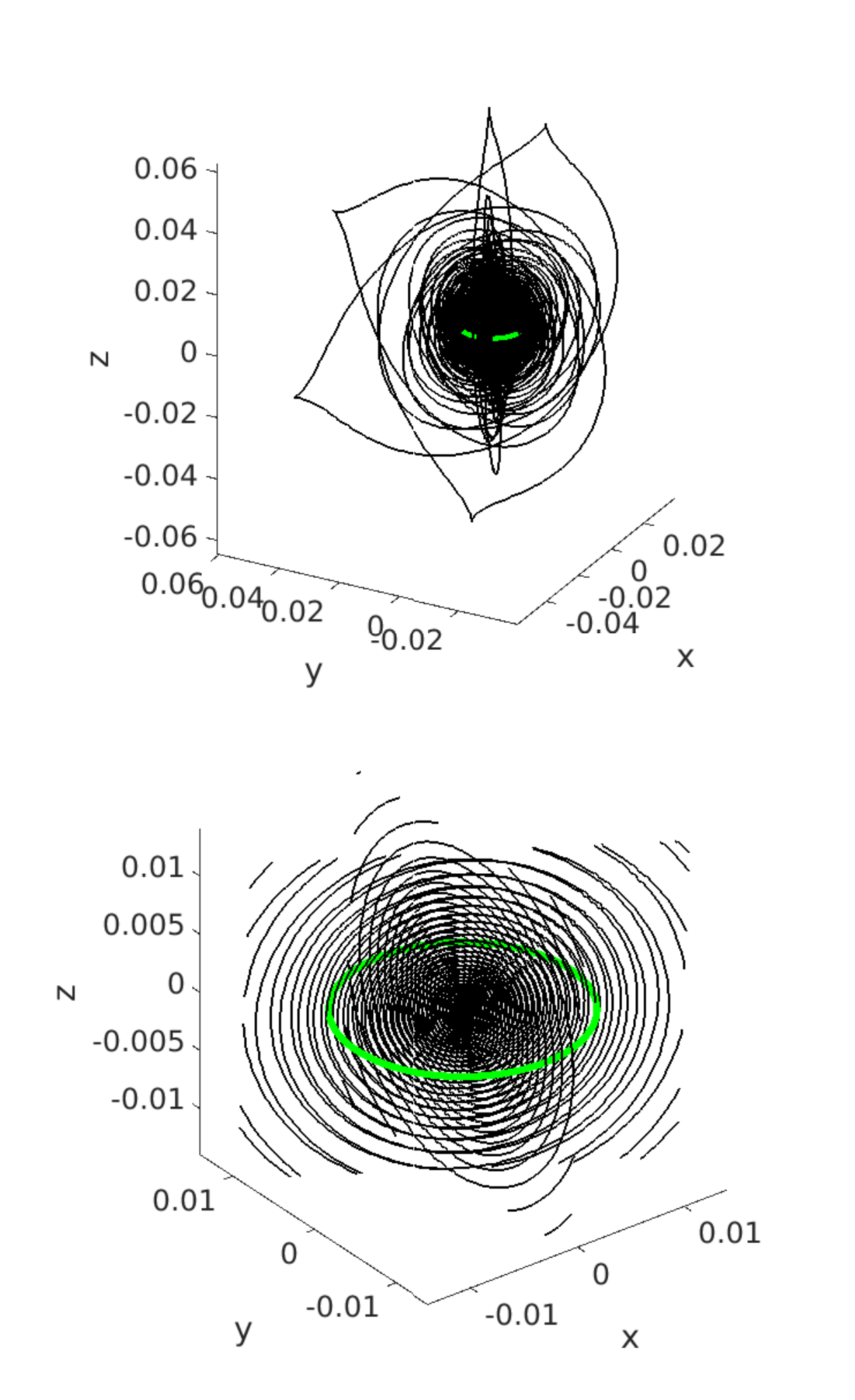}
  \caption{Example of a trajectory for a reflected ray with $\omega = 10\,\omega_\text{dyn}$ and $\ell = 1$ integrated up to $t = 400$ (in units of $\omega_\text{dyn}^{-1}$), with the radial extent of the field region ($R_f = 0.01$) indicated by the green circle. Here the central field strength is set to the critical value. The lower plot shows a zoom to near the centre, where for clarity the trajectory up to $t = 100$ only has been plotted.}
  \label{fig:trajectory}
\end{figure}

\begin{figure}
  \centering
  \includegraphics[width=\columnwidth]{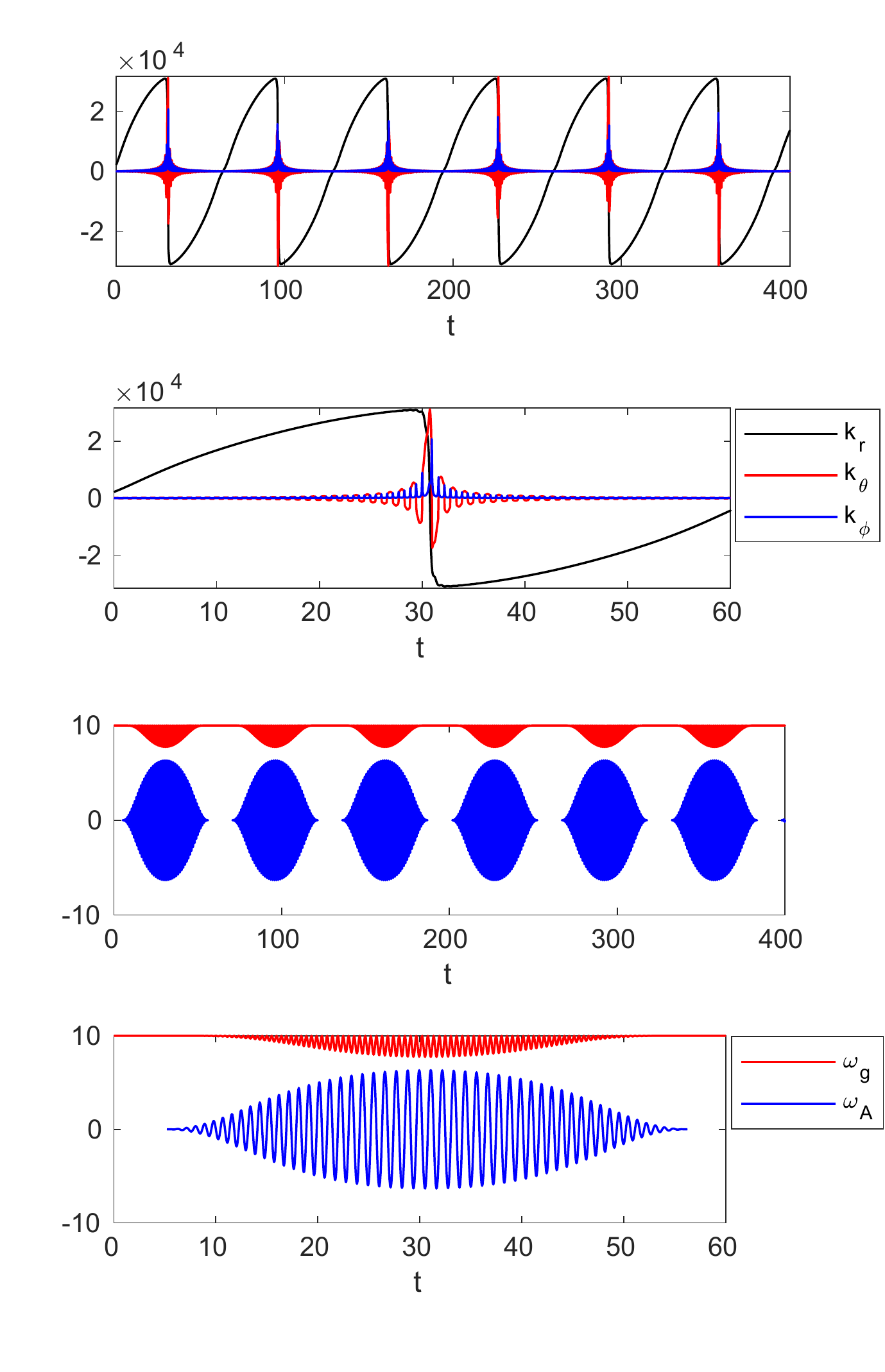}
  \caption{Top: components of the wavevector as a function of time, for the ray shown in Figure \ref{fig:trajectory}. The second panel shows a zoom into the interval $t \in [0,60]$, so that oscillatory features can be seen more clearly. The third panel shows the Alfv\'{e}n and gravity wave frequencies $\omega_A$ and $\omega_g$ versus time, and the fourth panel shows the same but restricted to the interval $t \in [0,60]$. Outside the field region $\omega_A$ has not been plotted, but its value there is zero.}
  \label{fig:k_om}
\end{figure}

\section{Methods}\label{sec:methods}
\subsection{Magneto-gravity ray tracing}\label{sec:raytracing}
Full details of the ray initialisation and integration procedure can be found in \citet{Loi2020}. To summarise, Hamiltonian ray tracing involves solving the initial value problem for the wave packet trajectories, described at any point in time by the position vector $\mathbf{x}$ and wavevector $\mathbf{k}$. Their time evolution is governed by Hamilton's equations
\begin{align}
  \frac{\rmd \mathbf{x}}{\rmd t} = \nabla_\mathbf{k} H \:,\quad \frac{\rmd \mathbf{k}}{\rmd t} = -\nabla H \:, \label{eq:Hamilton}
\end{align}
where the Hamiltonian is given by $H(\mathbf{x}, \mathbf{k}, t) = \omega(\mathbf{x}, \mathbf{k}, t)$. In the case of magneto-gravity waves, $\omega^2 = \omega_A^2 + \omega_g^2$, where
\begin{align}
  \omega_A &= \mathbf{k} \cdot \mathbf{v}_A \:, \label{eq:omA} \\
  \omega_g &= \frac{k_\perp}{k} N \label{eq:omg}
\end{align}
are the Alfv\'{e}n and gravity wave frequencies. Here $\mathbf{v}_A$ is the Alfv\'{e}n velocity, $k \equiv |\mathbf{k}| = \sqrt{k_r^2 + k_\perp^2}$ is the wavenumber and $k_\perp \equiv \sqrt{k_\theta^2 + k_\phi^2}$ is the horizontal component of the wavevector.

In three spatial dimensions, Hamilton's equations form a system of six linear ordinary differential equations which can be solved by standard techniques. The explicit forms of these equations in spherical polar coordinates are shown in \citet{Loi2020}, equations (23)--(28). In practice, they were solved using a fourth-order Runge-Kutta scheme with a time step size of 0.001 for a total duration of 200 time units (time is measured in units of $\omega_\text{dyn}^{-1}$). For each combination of parameters ($\omega$, $\ell$, $R_f$ and $v_{A,\text{cen}}$) 1200 rays were launched from $r > R_f$ (i.e.~outside the field region) into the magnetised core, with different latitudes and wavevector polarisations in order to sample the full sphere\footnote{In practice, all rays were launched from $r = 2R_f$, but the actual launch radius does not matter; this is discussed in more detail in \citet{Loi2020}, section 3.1.3.}. There were observed to be two possible outcomes for the rays, either \textit{trapping} or \textit{reflection}. Trapping is associated with a continuous divergence of the wavevector implying eventual dissipation, so such waves are unlikely to be involved in mode formation. However, mode formation may be possible using the reflected rays, which exhibit periodic bouncing between the inner and outer turning points of the g-mode cavity, akin to waves in the unmagnetised case. One might therefore expect EBK quantisation principles to be applicable to this subset of rays, to which we hereafter restrict our attention.

An example of a reflected trajectory for a run where $v_{A,\text{cen}} = v_{A,\text{crit}}$ is shown in Figure \ref{fig:trajectory}. This is seen to undergo six bounces over an interval of 400 time units. The time evolution of the wavevector components, as well as the Alfv\'{e}n and gravity wave frequencies, are shown in Figure \ref{fig:k_om}. A characteristic property of reflected rays is that the envelopes of $\omega_g$ and $\omega_A$ overlap minimally (this behaviour can be seen in the lower panels of Figure \ref{fig:k_om}), whereas for trapped rays they strongly overlap.

\begin{figure}
  \centering
  \includegraphics[width=\columnwidth]{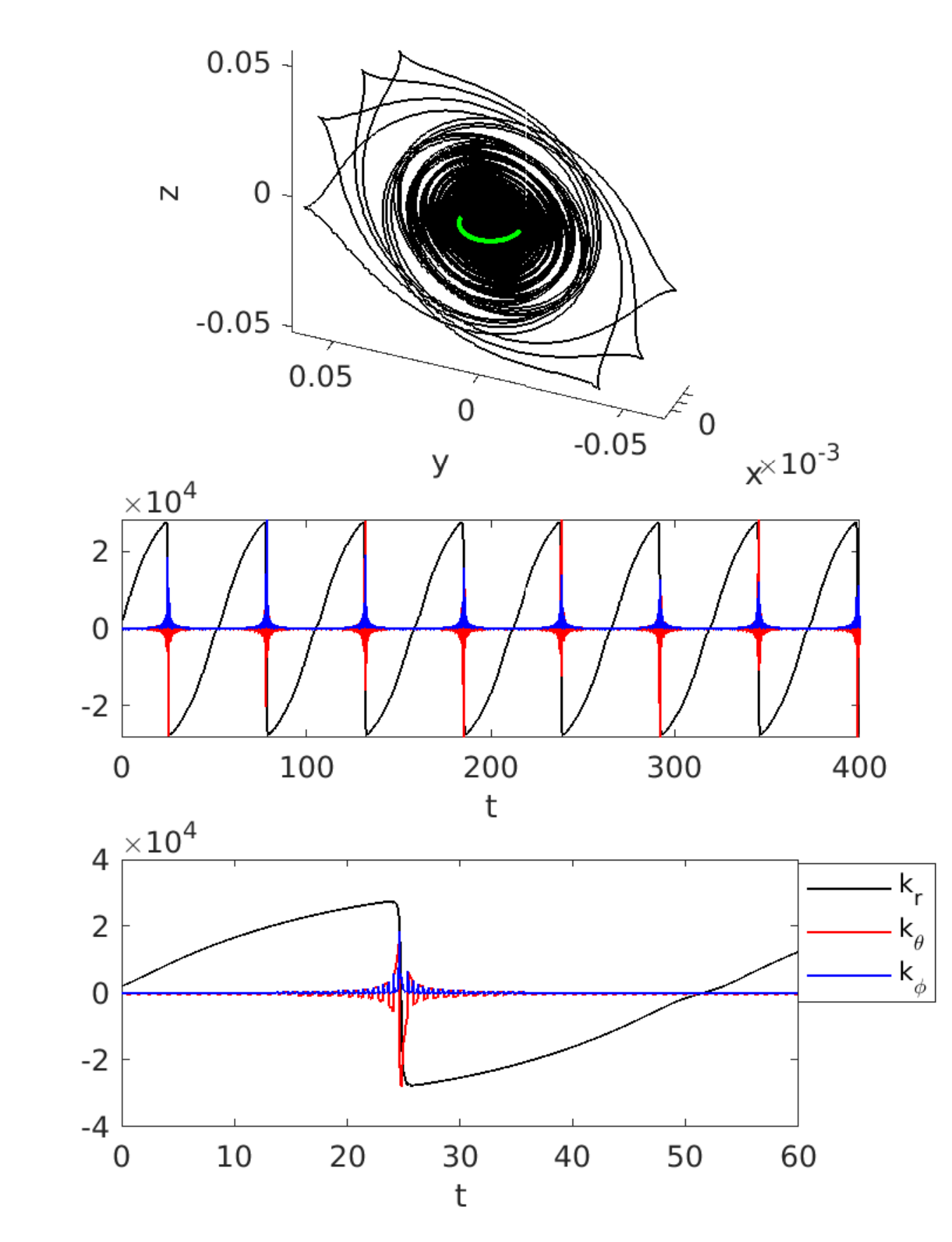}
  \caption{Top: the trajectory of a ray propagating under conditions of zero field, with the same launch latitude and wavevector polarisation as the ray in Figure \ref{fig:trajectory}. Middle and bottom: wavevector components as a function of time, where the bottom panel is a zoom in to the interval $t \in [0,60]$.}
  \label{fig:nofield}
\end{figure}

For comparison, the trajectory of the ray launched with the same initial parameters but into an unmagnetised core is shown in Figure \ref{fig:nofield}. Similar periodicities can be seen, with the main qualitative difference being that the trajectory in the unmagnetised case is confined to a plane (a known property of Hamiltonian trajectories in a spherically symmetric potential), whereas this is no longer true when the field is switched on. Since the magnetic field is axisymmetric, in both cases $L_z = k_\phi r \sin \theta$ is well conserved (to within one part in $10^3$), but only in the unmagnetised case is $L = k_\perp r$ conserved, since magnetic fields break spherical symmetry.

\subsection{Phase-space volume measurement}\label{sec:volume_measurement}
\begin{figure}
  \centering
  \includegraphics[width=\columnwidth]{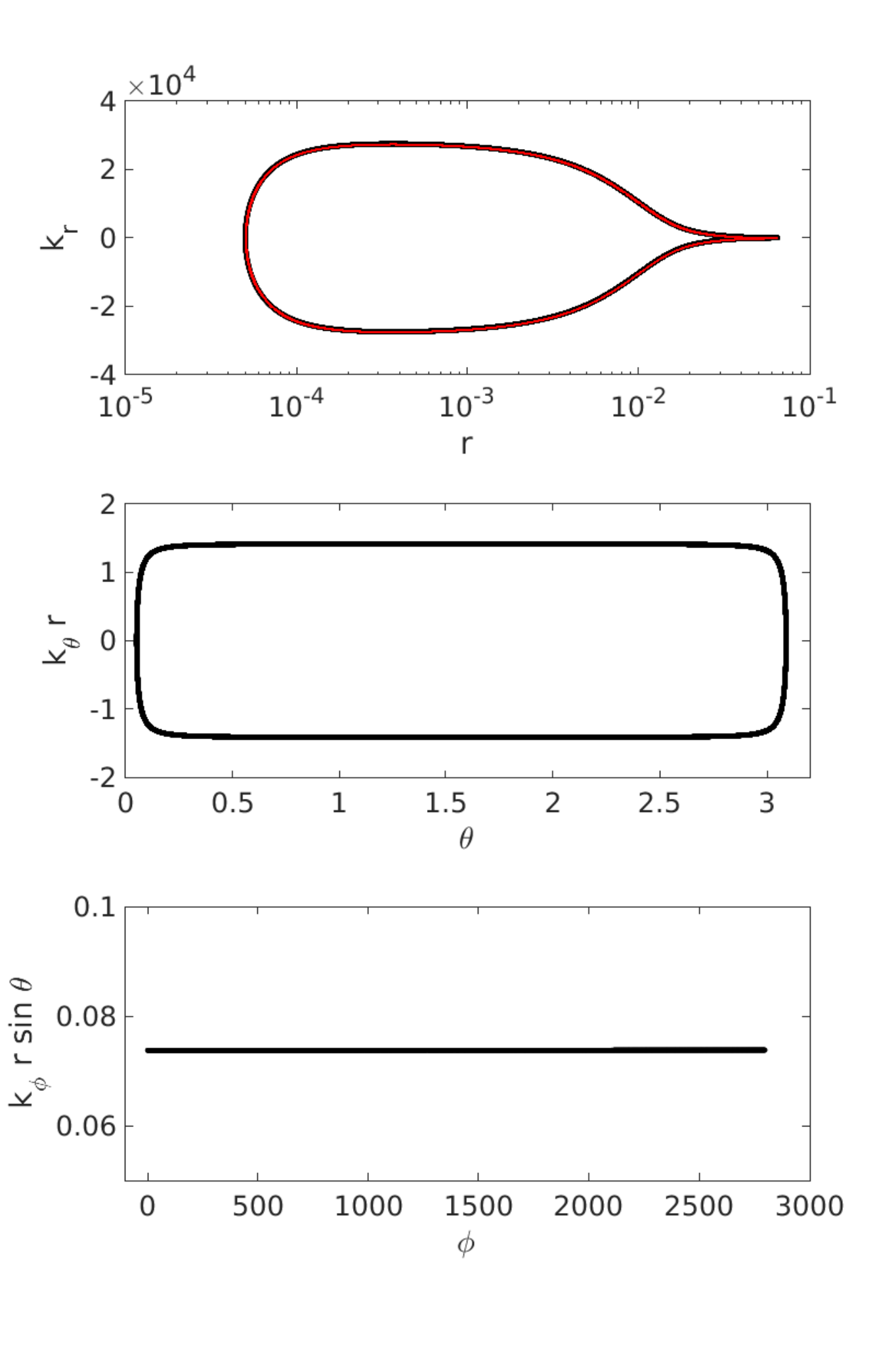}
  \caption{Action-angle diagrams plotting each conjugate momentum against the corresponding spatial coordinate, for the ray in Figure \ref{fig:nofield} ($\omega = 10\,\omega_\text{dyn}$, $\ell = 1$, $v_A = 0$). This trajectory forms closed orbits in action-angle space, and the enclosed areas in the top two panels correspond to the conserved actions associated with a fully integrable Hamiltonian system. In the bottom panel, the quantity on the vertical axis is $L_z$, the conserved action itself. The red line in the top panel corresponds to the theoretical prediction for $k_r$ as a function of $r$ based on the gravity-wave dispersion relation. The ray tracing data (underlying black points) are in close agreement with this. Note the logarithmic scale on the horizontal axis in the top panel. Although the different axes have different units and scalings, the areas of each subpanel all have the same units (dimensionless) so their numerical values can be directly compared. E.g.~in the top panel, $\int k_r \rmd r$ is a roughly triangular area of base length $4 \times 10^4$ and height 0.1, while in the middle panel, $\int k_\theta r \rmd \theta$ is an approximately rectangular area of width 2 and length 3.}
  \label{fig:pdq_nofield}
\end{figure}

\begin{figure}
  \centering
  \includegraphics[width=\columnwidth]{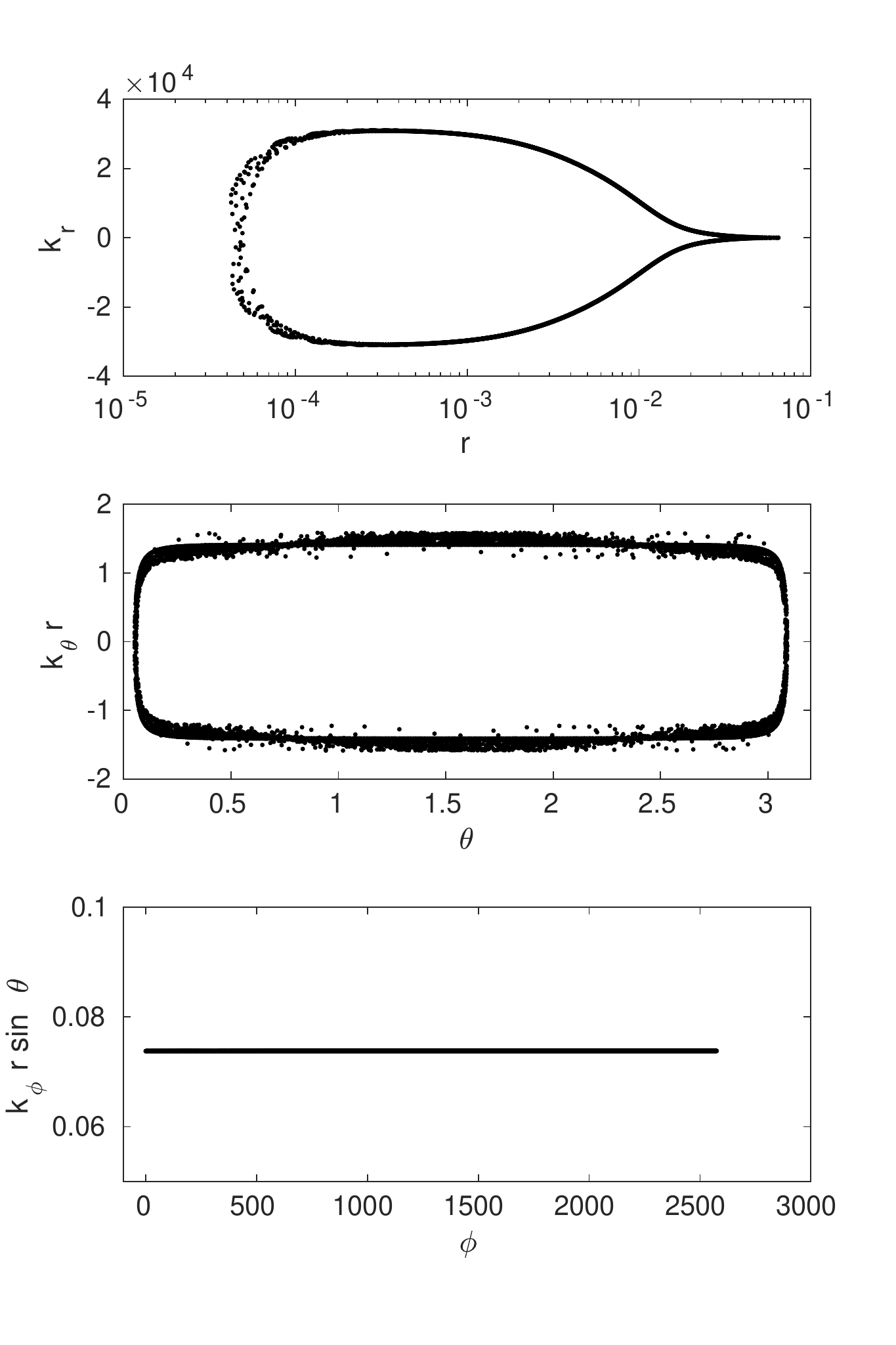}
  \caption{As for Figure \ref{fig:pdq_nofield}, but for a ray with $v_{A,\text{cen}} = v_{A,\text{crit}}$ (trajectory shown in Figure \ref{fig:trajectory}). In this case, the magnetic field has broken the spherical symmetry and so the curves in action-angle space are no longer perfectly closed.}
  \label{fig:pdq}
\end{figure}

As discussed in Section \ref{sec:intro}, the volume occupied by a region of phase space associated with a group of modes is roughly proportional to the maximum quantum number of that set of modes (number of modes that ``fit'' within that volume), as given in Equation (\ref{eq:nWeyl}). In three dimensions, this would be an integral over six-dimensional phase space. However, we notice that for both gravity and magneto-gravity modes, the volume associated with the radial degree of freedom dominates significantly over the angular degrees of freedom. This can be seen in Figures \ref{fig:pdq_nofield} and \ref{fig:pdq}, which show the curves of $p_i$ versus $q_i$, the enclosed areas of which correspond to the dimensionless phase-space volumes associated with each degree of freedom. In both cases, the top panel (radial degree of freedom) encloses about three orders of magnitude more area than the latitudinal degree of freedom (middle). The bottom panel shows the azimuthal degree of freedom, which due to axisymmetry is in fact $L_z$, the conserved action itself. Its value is also several orders of magnitude smaller than the area associated with the radial degree of freedom. We therefore choose to ignore the angular degrees of freedom and approximate the phase-space volume solely using the radial degree of freedom, i.e.~area of the curve in the $r$-$k_r$ plane.

Now the dispersion relation for a pure gravity wave is given by $\omega = \omega_g$, where $\omega_g$ is defined in (\ref{eq:omg}). If we consider fixed $\ell$, noting that $k_\perp = \sqrt{\ell(\ell+1)}/r$, then this can be rearranged into
\begin{align}
  k_r(r, \omega) = \frac{\sqrt{\ell(\ell+1)}}{r} \left( \frac{N^2}{\omega^2} - 1 \right)^{1/2} \:. \label{eq:kr_grav}
\end{align}
This theoretical relationship has been plotted as the red line in the top panel of Figure \ref{fig:pdq_nofield}, with which the ray tracing data (underlying black points) are well consistent. For an argument illustrating why the enclosed area should be the relevant quantity to consider, note that gravity waves cannot propagate if $\omega > N$, so effectively the g-mode spectrum for fixed horizontal wavenumber ``fills down'' starting from $\omega = N$ to smaller values of $\omega$, with higher (absolute\footnote{Although common convention to assign negative radial orders to g-modes in order to avoid clashing with p-mode indexing, in this work (since we do not deal with p-modes) we will regard g-mode orders as being positively signed.}) radial orders associated with smaller $\omega$. That is, the volume of phase space occupied by g-modes of radial order less than some reference value $n_\text{ref} \equiv n_W(\omega_\text{ref})$ corresponds to the region where $\omega_\text{ref} < \omega < N$. This can be equivalently stated as $|k_r| < k_{r,\text{ref}}$, where $k_{r,\text{ref}} \equiv |k_r(r, \omega_\text{ref})|$. Neglecting the other degrees of freedom, the number of modes predicted to fit into the associated phase-space volume given by the Weyl formula (\ref{eq:nWeyl}) is then
\begin{align}
  n_W(\omega_\text{ref}) &\approx \frac{1}{2\pi} \iint_{\omega < N} \rmd k_r \rmd r \nonumber \\
  &= \frac{1}{2\pi} \int_{r_1}^{r_2} \int_{-k_{r,\text{ref}}}^{k_{r,\text{ref}}} \rmd k_r \rmd r = \frac{1}{\pi} \int_{r_1}^{r_2} k_r(r, \omega_\text{ref}) \rmd r \label{eq:n}
\end{align}
which leads to the expression in (\ref{eq:ngrav}), matching the known asymptotic result for g-modes in Equation (\ref{eq:n_WKB}), as derived by \citet{Tassoul1980}.

Unfortunately, a similarly elegant expression cannot be obtained for magneto-gravity modes, since the presence of the Alfv\'{e}n term gives rise to the need to invert a quartic in $k_r$. However, it is still possible to compute the phase-space volume integral $\int k_r \rmd r$ numerically. It is to be noted from Figure \ref{fig:pdq} that in the magneto-gravity case the trajectories do not form perfectly closed curves, owing to the breaking of spherical symmetry and consequent non-integrability. However, it is remarkable how close they are to being so, which suggests that the dynamics for magneto-gravity waves may still be near-integrable even at high field strengths, and that the associated spectrum may be approximately regular. Since $k_r$ is reasonably well defined at each $r$, we choose to compute the enclosed area by dividing up the horizontal axis into a set of radial bins and then finding the median value of $k_r$ in that bin, over all reflected rays for a particular run (combination of $\omega$, $\ell$, $R_f$ and $v_{A,\text{cen}}$). The quantity $\int k_r \rmd r$ is then evaluated as a Riemann sum.

To obtain an estimate of errors, we considered two different sources: (i) the binning error, which was taken to be the standard deviation over five choices of bin sizes, dividing the full range of $r$ into either 200, 250, 300, 350 or 400 points, and (ii) the dispersion within the bins due to non-integrability, quantified through the interquartile range. In the latter case, the error was taken to be half the difference in volume computed using the set of upper quartiles, and the set of lower quartiles. This was done using the 300-bin division, but was observed to depend very little on the binning choice. In practice, the second source of error (intrinsic dispersion due to non-integrability) was observed to dominate over the binning error, and more so for larger field strengths. The overall error in volume measured at each $\omega$ was taken to be the quadrature sum of these two sources. The fractional error in $\rmd n_W/\rmd \omega$ was then taken to be the ratio of the frequency-averaged volume error to the dynamic range (i.e.~difference between maximum and minimum volumes measured over the full frequency range). The use of a quantile approach (median and interquartile range, rather than mean and standard deviation) to capture the characteristic value and dispersive error was motivated by the non-Gaussian nature of the data.

\section{Results}\label{sec:results}
\begin{figure}
  \centering
  \includegraphics[width=\columnwidth]{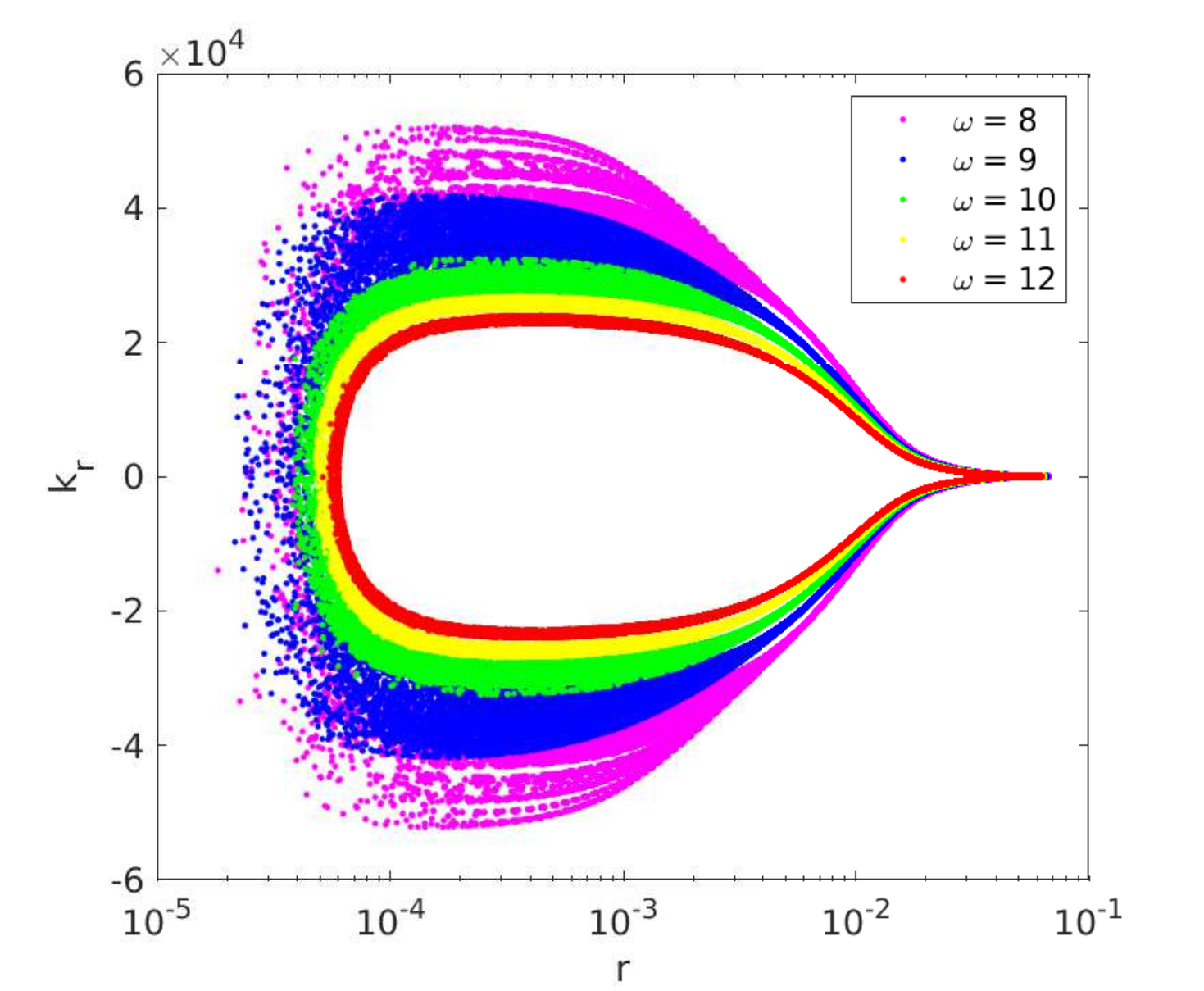}
  \caption{Phase-space plot of all reflected rays with $\ell = 1$ and $v_{A,\text{cen}} = v_{A,\text{crit}}$, where different frequencies are shown in different colours. Rays with higher frequencies are seen to enclose smaller areas. For clarity, only one point per 4 time units along each trajectory has been plotted.}
  \label{fig:kr_vs_r_om}
\end{figure}

For pure g-modes of fixed $\ell$ and high radial order, it is approximately true that $\omega = \omega_g \propto 1/k_r$, so as $\omega$ decreases the curves in the $r$-$k_r$ plane enclose larger and larger areas. Lower frequencies are thus associated with higher radial orders. However, the opposite dependence is expected for Alfv\'{e}n waves, since $\omega = \omega_A \propto k$. In the case of magneto-gravity waves, for which the dispersion relation $\omega^2 = \omega_A^2 + \omega_g^2$ is the quadrature sum of both terms, the direction of variation in $k_r$ with $\omega$ depends on which term is dominant. Reflected rays are those for which $\omega_g$ dominates over $\omega_A$, so it might be expected that their behaviour should be qualitatively similar to that of pure gravity rays. This is indeed the case, as can be seen in Figure \ref{fig:kr_vs_r_om}, which plots $k_r$ versus $r$ at different frequencies for a strong field ($v_{A,\text{cen}} = v_{A,\text{crit}}$), for $\ell = 1$. Curves associated with lower frequencies enclose larger areas than those of higher frequencies.

\begin{figure}
  \centering
  \includegraphics[width=\columnwidth]{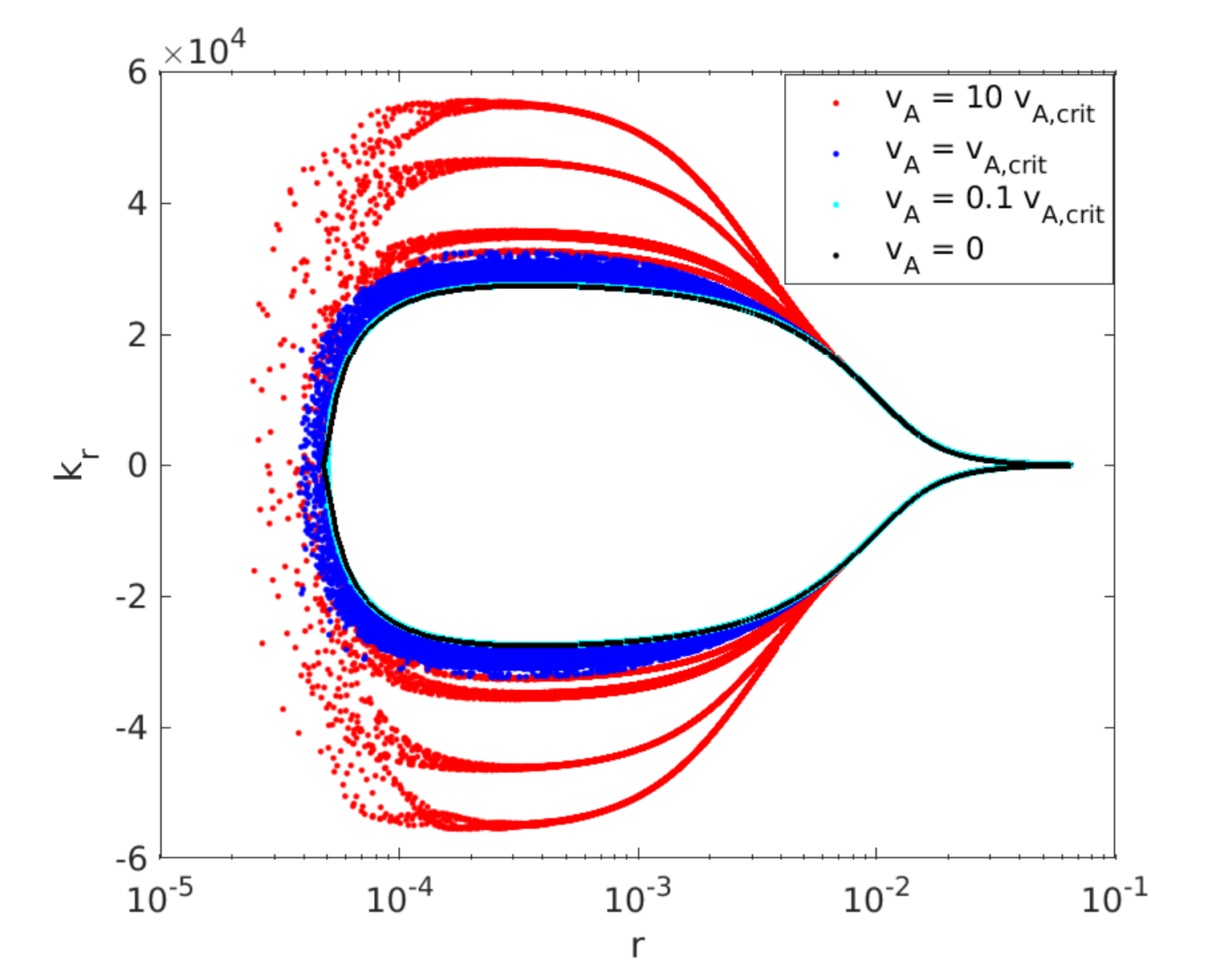}
  \caption{Phase-space plot for all reflected rays with $\omega = 10$ and $\ell = 1$, comparing zero (black), 0.1 times critical (cyan), critical (blue) and 10 times critical (red) field strengths. For clarity, only one point per 4 time units along each trajectory has been plotted. The discreteness in $k_r$ for the highest field strength is due to discreteness in the launch parameters, combined with the intrinsically large dispersion associated with non-integrability.}
  \label{fig:kr_vs_r_vA}
\end{figure}

\begin{figure}
  \centering
  \includegraphics[width=\columnwidth]{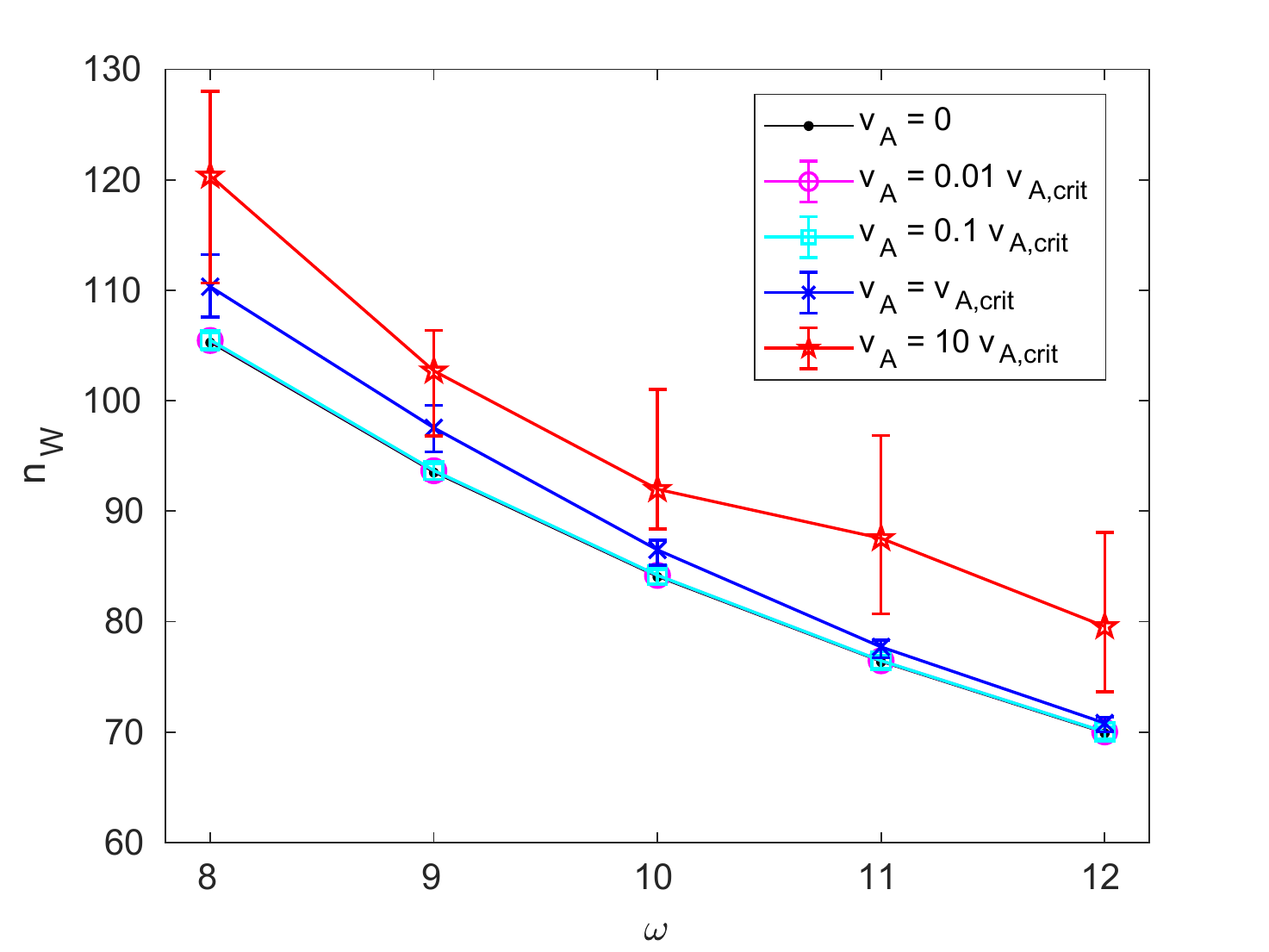}
  \caption{``Effective'' radial order (volume of phase space divided by $2\pi$) as a function of frequency $\omega$ for modes with $\ell = 1$. Different field strengths are compared, these shown in different colours as indicated in the figure legend.}
  \label{fig:n_vs_om_vA}
\end{figure}

\begin{figure}
  \centering
  \includegraphics[width=\columnwidth]{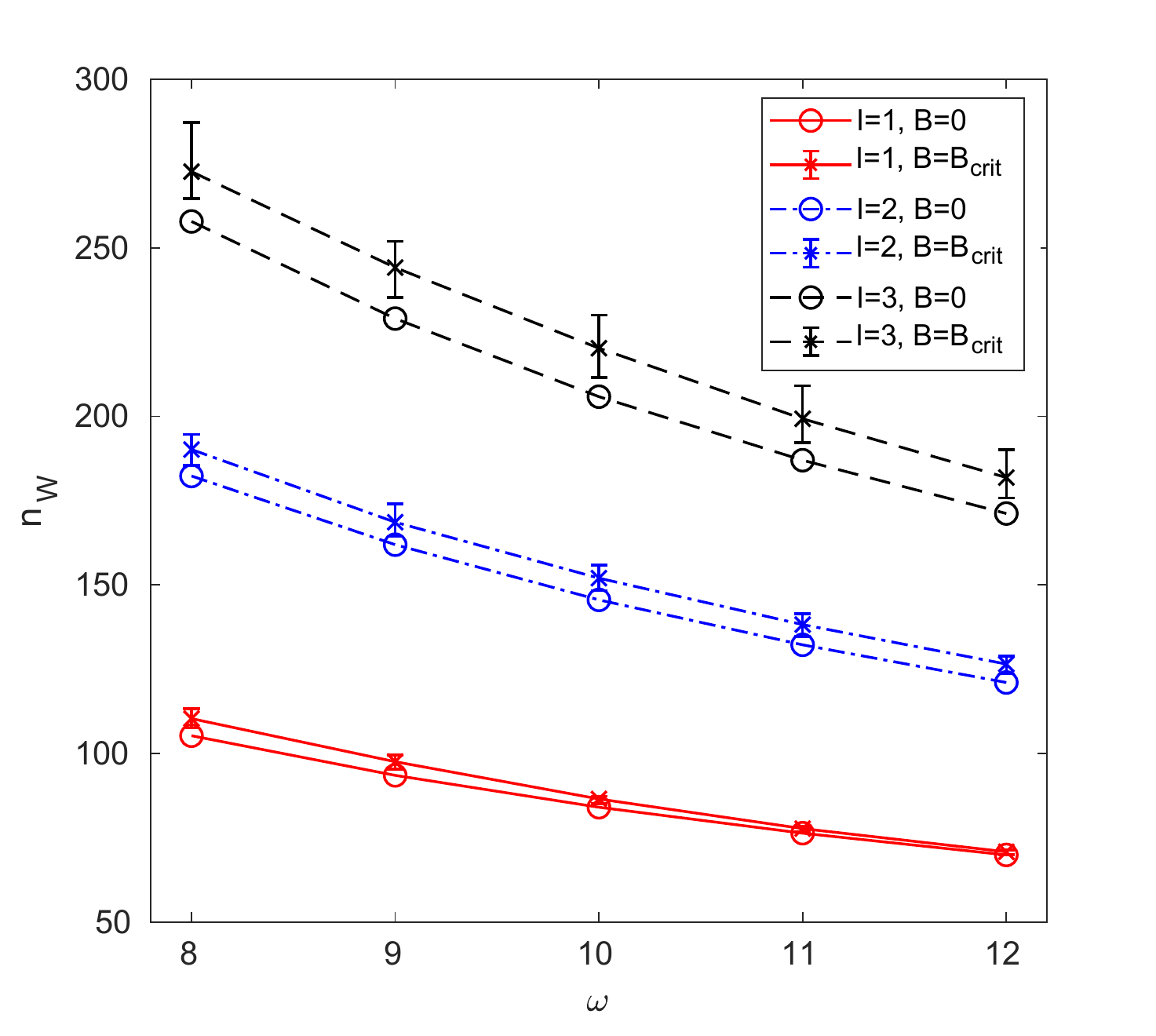}
  \caption{Effective radial order versus frequency, comparing zero field (circles) and critical field (crosses). Red solid, blue dash-dotted and black dashed lines correspond to $\ell = 1$, 2 and 3, respectively.}
  \label{fig:n_vs_om_l}
\end{figure}

As a comment, we also observe that the curves enclose larger areas when the field is switched on, compared to when it is zero. This can be attributed to the need for $k_r$ to be larger in order to form a mode with the same frequency as a pure g-mode, since $\omega_g$ needs to be smaller. This behaviour can be seen in Figure \ref{fig:kr_vs_r_vA}, which plots $k_r$ versus $r$ for different field strengths. This suggests that the presence of a magnetic field acts to increase the effective radial order $n_W$ at any given frequency. In Figure \ref{fig:n_vs_om_vA} we plot the effective radial order, given by the enclosed area divided by $2\pi$ as per Equation (\ref{eq:n}), as a function of $\omega$, at different field strengths, for $\ell = 1$. It can be seen that the field has minimal effect ($<$1\%) on $n_W$ until the central field strength passes the critical value. Around this point the increase in effective radial order becomes around 10\%. Increases of a similar magnitude are seen for $\ell = 2$ and 3, as shown in Figure \ref{fig:n_vs_om_l}. 

The quantity we wish to measure is the rate of change in enclosed area $\int k_r \rmd r$ with respect to $\omega$, since this is directly proportional to the number of modes per unit frequency interval, $\rmd n_W/\rmd \omega$. Comparing this to the same quantity in the absence of a field then determines how much the field changes the average spacing. As discussed above, it appears that the phase-space trajectories of magneto-gravity rays bound nearly closed curves, suggesting the possibility of pseudo-regular spectra. It may be, given their gravity-dominated character, that magneto-gravity modes formed from the constructive interference of reflected rays are roughly equally spaced in period akin to pure g-modes, but this period may incorporate Lorentz as well as buoyancy terms, and exhibit some variation with frequency. We shall therefore loosely use the term `period spacing' to generally mean `mode spacing', with the understanding that regularity in period space may not be exact. Note that the quantity measured here is simply the average density of modes. To measure this, we approximated $\rmd n_W/\rmd \omega$ for each frequency interval $\omega \in [8,9]$, $[9,10]$, $[10,11]$, $[11,12]$ by $\Delta n_W/\Delta \omega$ over that interval (see Figure \ref{fig:n_vs_om_vA}). The fractional difference in these $\rmd n_W/\rmd \omega$ values with and without a magnetic field was calculated within each interval, and then averaged over the four intervals. Also note that only propagation within the g-mode cavity is investigated; we do not examine coupling with p-modes, which should take place in identical fashion to the non-magnetic case and is a well studied process. In that sense, the results obtained here pertain to the magnetic equivalent of the asymptotic period spacing of pure g-modes, except that this is derived from Hamiltonian phase space considerations, rather than an asymptotic analysis.

Table \ref{tab:results} summarises the results. Strong magnetic fields are associated with a characteristic increase in density of modes (decrease in period spacing) of about 10\%, compared to the non-magnetic case. This has been measured for three different field region sizes $R_f$ and degrees $\ell$, with the estimated change in period spacing similar across this range of input parameters. At sub-critical field strengths, the measured change in period spacing is much smaller, consistent with zero within the measurement uncertainties. It would seem that the increase in density of modes is driven by the increase in effective radial order associated with the appearance of the Alfv\'{e}n term in the dispersion relation, which is positive definite and therefore at fixed $\omega$ necessitates a larger $k_r$, compared to no field. This boosts the volume of phase space available at any given frequency, leading to a proportional increase in the rate of change of this volume with respect to $\omega$. More modes are then able to ``fit'' within a given frequency interval, shrinking the period spacing.

\begin{table}
  \label{tab:results}
  \caption{Results for the change in density of modes over the interval $\omega \in [8,12]$ due to a magnetic field, comparing different field radii $R_f$, central Alfv\'{e}n speeds $v_{A,\text{cen}}$ and spherical harmonic degrees $\ell$. The rightmost column shows the percentage increase in density of modes compared to zero field. Errors correspond to the combined binning and dispersive error, quantified as described at the end of Section \ref{sec:volume_measurement}.}
  \centering
  \begin{tabular}{cccc}\hline
    $R_f/R_*$ & $v_{A,\text{cen}}/v_{A,\text{crit}}$ & Degree & $\Delta( \rmd n_W/\rmd \omega )$ (\%) \\ \hline
    0.006 & 1.0 & $\ell=1$& $9 \pm 3$ \\ \hline
    0.01 & 0.01 & $\ell=1$ & $0.2 \pm 1.9$ \\
    & 0.1 & $\ell=1$ & $0.3 \pm 1.9$ \\
    & 1.0 & $\ell=1$ & $12 \pm 4$ \\
    & & $\ell=2$ & $4 \pm 4$ \\
    & & $\ell=3$ & $6 \pm 5$ \\
    & 10 & $\ell=1$ & $11 \pm 17$ \\ \hline
    0.014 & 1.0 & $\ell=1$ & $14 \pm 4$ \\ \hline
  \end{tabular}
\end{table}

In light of these considerations, we suggest that it might not be physically possible for magneto-gravity modes to differ in period spacing from pure g-modes by more than several tens of per cent. This is because the transition from reflection to trapping occurs when $\omega_A$ grows large enough to begin to overlap with $\omega_g$ in value, and trapped rays are not expected to be involved in mode formation because they experience total dissipation. Hence the characteristic contribution of $\omega_A$ to the total $\omega$ for reflected rays would not be able to exceed several tens of per cent, which sets a corresponding upper limit on the increase in effective radial order due to a magnetic field. This in turn appears to be a reasonable proxy for the fractional increase in $\rmd n_W/\rmd \omega$, which consequently is also likely to be upper-bounded by several tens of per cent compared to no field. Our measured value of $\sim$10\% is consistent with this notion. Note that the exact field strength at which a ray transitions from being reflected to trapped depends on its particular launch parameters (location and wavevector), so for a large ensemble of rays with different starting parameters, the overall rate of trapping undergoes a gradual transition from zero as the field strength is increased. It is thought that this should asymptotically approach unity for large field strengths, although numerical issues make this difficult to directly verify (see discussion in Section \ref{sec:limitations}).

\section{Discussion}\label{sec:discussion}
\subsection{Comparison with observations}
In red giant stars, g-modes localised to the core can be observed via their coupling to a p-mode of nearby frequency, which forms mixed modes with detectable surface amplitudes. The spectrum of g-modes is much denser than p-modes, giving rise to multiple g-dominated mixed modes for every p-dominated mixed mode. If observations are of sufficient quality that enough g-dominated mixed modes can be detected, then a fit can be made to measure the asymptotic period spacing after suitably stretching to account for frequency-dependent shifts associated with the coupling process \citep{Vrard2016}. This has been done for stars with depressed dipole modes and it was found that the spacings did not significantly differ from normal stars \citep{Mosser2017}. Therein, the authors reported that measured period spacings were slightly lower than average, but claimed that this was in agreement with the higher mass distribution of depressed stars. While they definitively ruled out an increase in period spacings, it is unclear whether a systematic decrease was similarly ruled out. Our prediction of $\sim$10\% smaller period spacings may be consistent if we note that a major source of error in the parameter-fitting process comes from aliasing issues. These are of the order 10\% for typical stars \citep[][table 1]{Vrard2016} and are worse for stars with fewer detected mixed modes, which would pertain particularly to those stars with depressed modes.

Measurements of the rotational splitting in red giants with depressed dipole modes also found no significant difference in the distribution of these values, compared to the rest of the population \citep{Mosser2017}. In the absence of rotation, a magnetic field causes splitting into a multiplet of $\ell+1$ modes, with those of the same $|m|$ remaining degenerate \citep{Unno1989}. The combined presence of rotation and magnetism, where the rotation and magnetic axes may be misaligned, should give rise to $(2\ell+1)^2$ peaks in the spectrum for fixed $(n,\ell)$ when viewed in an inertial frame \citep{Dicke1982, Dziembowski1984, Gough1990}. In this situation, $m$ is no longer a well-defined quantum number, since there is no axis with respect to which each eigenfunction can be described by a single $m$. Rather, each of the $2\ell+1$ eigenfunctions becomes a linear combination of $2\ell+1$ basis functions, one for each $m$ defined with respect to a reference axis. One might expect that such a large increase in multiplicity (total of 9 modes for $\ell = 1$) should be possible to detect, especially since Lorentz splittings are likely to exceed rotational splittings if the field is strong. However, if strong magnetic fields are indeed the cause of mode depression, then the lack of such an occurrence may be linked to the orientation-selective nature of the trapping phenomenon. It was found by \citet{Loi2020} that the rays able to escape trapping and consequent dissipation corresponded to those orbiting near the magnetic equator with predominantly zonal polarisation vectors. These rays would be associated with sectoral modes. For $\ell = 1$, it may be that a strong magnetic field selectively damps the $|m| = 0$ mode (defined with respect to the magnetic axis), leaving only the $|m| = 1$ peak to be rotationally split. The number of detectable modes in each multiplet may then be reduced to values more similar to the rest of the population. This point remains speculative.

\subsection{Limitations}\label{sec:limitations}
It has to be emphasised that the measurement made in this work is an estimate of only the leading-order term for the change in period spacing due to a magnetic field. This was done using phase-space volume considerations, which work best in the limit of large radial orders and provide only a first approximation to the density of modes. The result quoted here is also an average over a small frequency range, whereas the full effects of a magnetic field would almost certainly be frequency dependent. Furthermore, the symmetry-broken nature of the problem would suggest that the mode spectrum should be more complicated than the non-magnetic case, with irregular subsets of modes present alongside regular/pseudo-regular ones. The present study has neither found nor characterised any irregular subsets, but focuses only on a pseudo-regular subset associated with the periodic reflected rays. It is thought that this should be the most easily detectable type of mode, and hence the most relevant for seismic diagnosis; however, characterisation of the full magneto-gravity mode spectrum remains incomplete.

In addition, it is to be noted that the arguments of this work are based entirely on Hamiltonian ray theory, and we have not directly solved for the mode spectrum. This would require full wave/eigenmode calculations in a 3D geometry, which are significantly more computationally expensive and complex to implement. Such calculations are likely to be the only way to accurately quantify all the frequency dependent effects of a strong magnetic field, and identify additional classes of modes. However, it is hoped that the results obtained here for pseudo-regular modes may still be useful for comparing to observations of actual period spacings.

Difficulties surround the extension of the approach used in this work to higher field strengths. The highest field strength considered here is 10 times critical, for which it can be seen that the errors are rather large (Table \ref{tab:results}). There are several reasons for this:
\begin{enumerate}
  \item The intrinsic dispersion associated with non-integrability drastically broadens as the field strength increases beyond critical (this can be seen from Figure \ref{fig:kr_vs_r_vA}), and so the characteristic value of $k_r$ at any $r$ becomes less well defined. The large uncertainties on the result for the highest field strength arise primarily from this effect.
  \item There is a shrinking of parameter space occupied by the periodic reflected rays, meaning that out of the original 1200 rays launched, as the field strength is increased, fewer and fewer are available for the measurement. This is the reason for the multiple tracks seen in the red points in Figure \ref{fig:kr_vs_r_vA}, which reflects the discreteness in launch parameters (latitude and wavevector polarisation).
  \item Systems exhibiting chaotic dynamics are highly susceptible to numerical errors, which grow more rapidly for more chaotic trajectories. The solutions obtained thus become progressively less reliable at higher field strengths: this can be inferred from the increased fluctuation of quantities that should otherwise be constants of motion (e.g.~$\omega$, $L_z$). Although not shown plotted here, at 100 times critical strength it can be seen that $\omega$ drifts by up to several tens of per cent within the duration of a single bounce.
\end{enumerate}
While the second source of error could be resolved through denser sampling, the first and third are intrinsic to the dynamics. Of these, the first issue dominates, reinforcing the point that the measurements here are inherently approximate, since there is in fact no ``true'' characteristic value for a non-integrable system.

General limitations of the ray tracing approach are also worth iterating: while advantageous over other methods in its computational speed and simplicity, this technique is formally exact in the short-wavelength limit and only incorporates WKBJ to zeroth order. This means that higher-order WKBJ effects such as amplitude growth and decay, phase dependencies, tunnelling, and partial transmission cannot be captured. However, it is not expected that any of these should drastically impact the phase-space volume measurement, which just requires accurate evolution of the wavevector. The observation that quantities such as $L$ and $L_z$ are well conserved, and that the theoretical and ray-traced values of $k_r$ as a function of $r$ are in close agreement (Figure \ref{fig:pdq_nofield}, top), suggest this to be the case. The accuracy of our result is likely to be more heavily limited by the considerations in preceding paragraphs.

\subsection{Related studies}
There has previously been, and continues to be, ongoing interest in the effects of a magnetic field on the seismic parameters of g-modes \citep{Hasan2005, Prat2019, Prat2020, VanBeeck2020}. So far, these related studies have focused on the case where magnetic fields are weak enough to be amenable to a perturbative treatment, and the desired quantity to be examined is the Lorentz splitting within a given multiplet. In the recent papers by \citet{Prat2019, Prat2020} which incorporated rapid rotation, they found that it should be possible to detect a weak magnetic field via the distortions it induces in the g-mode period spacing pattern. The work of \citet{Loi2019} attempted to extend analytic methods up to moderate field strengths through an alternate ``non-perturbative'' formalism that remains valid for frequency shifts that are comparable to mode spacings. However, rotation was not incorporated, and neither is it here. The main difference between the current and aforementioned works is that we seek here to understand the more challenging regime of dynamically significant magnetic fields, these being too strong to be treated perturbatively. Our results suggest that further to the known effects of weak fields, which include a lifting of the $m$-degeneracy and localised distortions to the period spacing pattern, strong fields may produce more drastic, global changes to the seismic properties. These include an overall decrease in period spacing by $\sim$10\% due to the increase in density of modes, and the selective damping of modes with low $|m|$ as a result of the trapping phenomenon.

As already noted in Section \ref{sec:limitations}, the ray theory employed here does not directly solve for the mode spectrum, but rather investigates local wave dynamics. Phase-space volume arguments are then invoked to infer properties of the global modes that may arise from constructive interference of those waves. Previous ray-based studies of magneto-gravity waves have similarly focused on local dynamics \citep[e.g.][]{Loi2018, Valade2018, Loi2020}; a formalism has yet to be developed that directly connects the results of ray theory to a quantisation condition for global magneto-gravity modes. Successes along this line have already been made in the context of inertial-acoustic \citep{Pasek2011, Pasek2012} and gravito-inertial \citep{Prat2017} modes in rapidly rotating stars, through clever manipulation of the dispersion relations and then application of EBK quantisation to derive analytic conditions for mode formation. However, the dispersion relations incorporating only rotation are easier to manipulate than those incorporating magnetism, since without a field they are only quadratic in the wavevector, whereas the magneto-gravity dispersion relation is quartic in the wavevector. Some progress with this may be possible, but it is beyond the scope of the current work.

\section{Summary}\label{sec:summary}
Using Hamiltonian ray calculations and phase-space volume arguments, we have estimated the average density of gravity modes in the presence of a strong magnetic field, and compared this to the situation without a field. We infer that strong fields result in an increase of $\sim$10\% in the density of modes, translating to a $\sim$10\% decrease in asymptotic period spacings, compared to no field. Neither the radius of the field region nor the spherical harmonic degree were found to significantly influence this result. The increase in density of modes only occurs when field strengths surpass the critical value; an order of magnitude below this, mode densities were unchanged compared to zero field, within measurement error. 

Physically, the increase in density of modes arises from an increase in effective radial order, which in turn is driven by the presence of the Alfv\'{e}n term in the magneto-gravity dispersion relation. This term is positive definite, implying a systematic decrease in period spacings. While some uncertainties lie in the values of $\sim$10\% measured here, we can still infer on the basis that mode-forming rays need to avoid becoming trapped that the decrease in period spacings cannot exceed several tens of per cent. Waves that become much more Alfv\'{e}nic than this are expected to be dissipated deep in the core, precluding their involvement in mode formation. The near integrability seen for reflected rays suggests that these are likely to form pseudo-regular spectra (perhaps spaced by a magnetically-modified period) that could be amenable to quantisation via the EBK method. The possibility of this, and/or the inclusion of rotation, may form the subject of future work.

\section*{Acknowledgements}
We thank John Papaloizou for useful discussions, and the anonymous reviewer for helpful comments and feedback. STL is supported by funding from Churchill College, Cambridge through a Junior Research Fellowship.

\section*{Data availability}
The data underlying this article will be shared on reasonable request to the corresponding author.

%%%%%%%%%%%%%%%%%%%%%%%%%%%%%%%%%%%%%%%%%%%%%%%%%%



\bibliographystyle{mnras}
%\bibliography{C:/Users/STCLoi/Documents/articles_and_papers/my_stuff/refs}

%%%%%%%%%%%%%%%%%%%%%%%%%%%%%%%%%%%%%%%%%%%%%%%%%%

%%%%%%%%%%%%%%%%% APPENDICES %%%%%%%%%%%%%%%%%%%%%

\appendix

\section{Radial flux function and derivative}\label{sec:appendix}

As mentioned in Section \ref{sec:magneticfield}, the choice of a decaying exponential density profile is convenient for allowing the twisted-torus field configuration to be written down in closed form, since the integrals in (\ref{eq:Psi}) can be performed analytically. By eliminating the need to evaluate the integrals numerically, one avoids the use of lookup tables/interpolation, which greatly reduces numerical errors associated with background non-smoothness when calculating the ray trajectories \citep[cf.~discussion in][section 3.1.2]{Loi2020}. Below are the resulting expressions for $\Psi(r)$ and $\Psi'(r)$, in terms of which the field components are defined according to Equations (\ref{eq:Br})--(\ref{eq:Bphi}). Here prime denotes a derivative with respect to the argument.

\begin{align}
  \Psi(r) &= \beta \lambda r \left\{ j_1(\lambda r) \left[ I_2(R_f) - I_2(r) \right] - y_1(\lambda r) \left[ I_1(R_f) - I_1(r) \right] \right\} \:, \\
  \Psi'(r) &= \frac{\Psi}{r} + \beta \lambda^2 r \left\{ j_1'(\lambda r) \left[ I_2(R_f) - I_2(r) \right] - y_1'(\lambda r) \left[ I_1(R_f) - I_1(r) \right] \right\} \nonumber \\
  &\quad - \beta \lambda r \left\{ j_1(\lambda r) I_2'(r) - y_1(\lambda r) I_1'(r) \right\} \:,
\end{align}
where
\begin{align}
  I_1(r) &\equiv \frac{\sigma \rme^{-r/\sigma}}{\lambda^2 (\sigma^2 \lambda^2 + 1)^3} \left\{ \left[ \lambda (\sigma^2 \lambda^2 + 1)^2 r^2 \right. \right. \nonumber \\
    &\quad \left. + \lambda \sigma (1 - 3 \sigma^2 \lambda^2) (\sigma^2 \lambda^2 + 1) r - 8 \sigma^4 \lambda^3 \right] \cos(\lambda r) \nonumber \\
  &\quad - \left[ \sigma \lambda^2 (\sigma^2 \lambda^2 + 1)^2 r^2 + (5\sigma^2 \lambda^2 + 1)(\sigma^2 \lambda^2 + 1) r \right. \nonumber \\
    &\quad \left. \left. - 2 \sigma^3 \lambda^2 (\sigma^2 \lambda^2 - 3) + \sigma (1 - \sigma^4 \lambda^4) \right] \sin(\lambda r) \right\} \:,
\end{align}
\begin{align}
  I_2(r) &\equiv \frac{\sigma \rme^{-r/\sigma}}{\lambda^2 (\sigma^2 \lambda^2 + 1)^3} \left\{ \left[ \sigma \lambda^2 (\sigma^2 \lambda^2 + 1)^2 r^2 \right. \right. \nonumber \\
    &\quad \left. (\sigma^2 \lambda^2 + 1)(5\sigma^2 \lambda^2 + 1) r + \sigma - 3\sigma^3 \lambda^2 (\sigma^2 \lambda^2 - 2) \right] \cos(\lambda r) \nonumber \\
  &\quad + \left[ \lambda (\sigma^2 \lambda^2 + 1)^2 r^2 + \sigma \lambda (\sigma^2 \lambda^2 + 1)(1 - 3\sigma^2 \lambda^2) r \right. \nonumber \\
    &\quad \left. \left. - 8 \sigma^4 \lambda^3 \right] \sin(\lambda r) \right\} \:,
\end{align}
\begin{align}
  I_1'(r) &= -\frac{1}{\sigma} I_1(r) - \frac{\sigma \rme^{-r/\sigma}}{\lambda^2 (\sigma^2 \lambda^2 + 1)^3} \left\{ \left[ \sigma \lambda^3 (\sigma^2 \lambda^2 + 1)^2 r^2 \right. \right. \nonumber \\
    &\quad \left. + \lambda (\sigma^2 \lambda^2 + 1)(3\sigma^2 \lambda^2 - 1) r + 8\lambda^3 \sigma^3 \right] \cos(\lambda r) \nonumber \\
  &\quad + \left[ \lambda^2 (\sigma^2 \lambda^2 + 1)^2 r^2 + \lambda^2 \sigma (\sigma^2 \lambda^2 + 1)(3 - \sigma^2 \lambda^2) r \right. \nonumber \\
    &\quad \left. \left. - 3\sigma^4 \lambda^4 + 6\sigma^2 \lambda^2 + 1 \right] \sin(\lambda r) \right\} \:,
\end{align}
\begin{align}
  I_2'(r) &= -\frac{1}{\sigma} I_2(r) + \frac{\sigma \rme^{-r/\sigma}}{\lambda^2 (\sigma^2 \lambda^2 + 1)^3} \left\{ \left[ \lambda^2 (\sigma^2 \lambda^2 + 1)^2 r^2 \right. \right. \nonumber \\
    &\quad \left. + \sigma \lambda^2 (\sigma^2 \lambda^2 + 1)(3 - \sigma^2 \lambda^2) r - 3 \sigma^4 \lambda^4 + 6 \sigma^2 \lambda^2 + 1 \right] \nonumber \\
  &\quad - \left[ \sigma \lambda^3 (\sigma^2 \lambda^2 + 1)^2 r^2 - \lambda (\sigma^2 \lambda^2 + 1)(1 - 3\sigma^2 \lambda^2) r \right. \nonumber \\
    &\quad \left. \left. + 8 \sigma^3 \lambda^3 \right] \sin(\lambda r) \right\} \:.
\end{align}

%%%%%%%%%%%%%%%%%%%%%%%%%%%%%%%%%%%%%%%%%%%%%%%%%%

% Don't change these lines
\bsp	% typesetting comment
\label{lastpage}
\end{document}